\documentclass[useAMS,usenatbib]{mn2e}
\usepackage{graphicx}
\usepackage{natbib}
\usepackage{Times}
\usepackage{enumerate}
\usepackage[fleqn]{amsmath}
\newcommand\lsim{~\lower.5ex\hbox{$\buildrel < \over \sim$}~}
\newcommand\gsim{~\lower.5ex\hbox{$\buildrel > \over \sim$}~}

\title[Non-axisymmetric perturbations in magnetized discs]{Revisiting linear
dynamics of non-axisymmetric perturbations in weakly magnetized accretion discs}

\author[G.R. Mamatsashvili et al.] {G.~R.~Mamatsashvili$^{1}$\thanks{E-mail:
george.mamatsashvili@tsu.ge},  G.~D.~Chagelishvili$^{2,3}$,
G.~Bodo$^{4}$ and P. Rossi$^{4}$\\
$^{1}$ Department of Physics, Faculty of Exact and Natural Sciences,
Tbilisi State
University, Il. Chavchavadze ave. 3, Tbilisi 0179, Georgia\\
$^{2}$ Abastumani Astrophysical Observatory, Ilia State University,
2a Kazbegi Ave., Tbilisi 0160, Georgia\\
$^{3}$ M. Nodia Institute of Geophysics, Tbilisi State University, Tbilisi 0193, Georgia\\
$^{4}$ INAF, Osservatorio Astronomico di Torino, Strada Osservatorio
20, Pino Torinese 10025, Italy}
\begin{document}

\date{Accepted 2013 August 4.  Received 2013 July 28; in original form 2013 May 25}

\pagerange{\pageref{firstpage}--\pageref{lastpage}} \pubyear{2013}

\maketitle

\label{firstpage}

\begin{abstract}
We investigate the linear dynamics of non-axisymmetric perturbations
in incompressible, vertically stratified Keplerian discs threaded by
a weak non-zero net vertical magnetic field in the local shearing
box approximation. Perturbations are decomposed into shearing waves,
or spatial harmonics whose temporal evolution is then followed via
numerical integration of the linearized ideal MHD equations of the
shearing box. There are two basic modes in the system --
inertia-gravity waves and magnetic mode, which displays the
magnetorotational instability (MRI). As distinct from previous
related studies, we introduce ``eigen-variables'' characterizing
each (counter-propagating) component of the inertia-gravity and
magnetic modes, which are governed by a set of four first order
\emph{coupled} ordinary differential equations. This allowed us to
identify a new process of linear coupling of the two above
non-axisymmetric modes due to the disc's differential rotation. We
also carried out a comparative analysis of the dynamics of
non-axisymmetric and axisymmetric magnetic mode perturbations. It is
demonstrated that the growth of ``optimal'' and close-to-optimal
non-axisymmetric harmonics of this mode, having transient nature,
can prevail over the exponential growth of axisymmetric ones (i.e.,
over the axisymmetric MRI) during dynamical time. A possible
implication of this result for axisymmetric channel solutions
emerging in numerical simulations is discussed. In particular, the
formation of the (axisymmetric) channel may be affected/impeded by
non-axisymmetric modes already at the early linear stage leading to
its untimely disruption -- the outcome strongly depends on the
amplitude and spectrum of initial perturbation. So, this competition
may result in an uncertainty in the magnetic mode's non-linear
dynamics. Although we consider incompressible perturbations, in the
final part, speculate on the dynamics in the compressible case. It
is shown that a maximum growth of non-axisymmetric magnetic mode
occurs at vertical wavelengths close to the disc scale-height for
which compressibility effects are important. This indirectly
suggests that compressibility plays a role in the dynamics of the
non-axisymmetric MRI and, ultimately, in the resulting turbulent
state.

\end{abstract}


\section{Introduction}

The interplay of magnetic field and differential rotation in
accretion discs offers a realistic hope of explaining energetic
processes -- turbulence and angular momentum transport -- in this
kind of astrophysical flows. Despite more than twenty years of
research in this field, after the seminal paper by
\citet{Balbus_Hawley91} in which the magnetorotational instability
(MRI) was proposed as the key element for angular momentum transport
in accretion discs, many questions are still open. A fully
consistent dynamical picture is still missing, mainly because of the
complex nature of the ingredients entering into dynamical phenomena
in magnetized discs.

In MRI studies, it is customary to consider two different kinds of
configurations: one in which there is a non-zero net magnetic flux
threading the disc and the second in which there is no net flux.  In
these two cases open problems and dynamical processes are somewhat
different, in fact, in the first case we have a well defined linear
instability, while in the second case the MRI must set in as a
subcritical instability and dynamo action must be effective in order
to sustain the field needed for the MRI to operate. In this paper,
we will focus our attention on the net flux case: in this case in
order to understand the dynamics one has to  consider the relative
importance and the  interplay of several different processes:
\begin{enumerate}[(I)]
\item
exponential instability of axisymmetric perturbations
\citep[axisymmetric magnetorotational
instability,][]{Balbus_Hawley91,Goodman_Xu94};
\item
transient growth of non-axisymmetric perturbations caused by the
shear/non-normality of the disc's differential rotation
\citep[][transient dynamics includes: Coriolis forces, the basic
disc flow shear, magnetic tension forces]{Balbus_Hawley92,
Tagger_etal92, Brandenburg_Dintrans06, Johnson07, Pessah_Chan12};
\item
non-linear interaction of axisymmetric perturbations (channel flow)
with non-axisymmetric (parasitic) ones
\citep{Goodman_Xu94,Pessah_Chan08,
Latter_etal09,Pessah_Goodman09,Longaretti_Lesur10,Pessah10};
\item
non-linear interaction between non-axisymmetric perturbations and
the onset and character of turbulence \citep[e.g.,][]{Hawley_etal95,
Brandenburg_etal95, Fromang_Papaloizou07, Lesur_Longaretti11,
Fromang_etal07, Bodo_etal08, Guan_etal09, Davis_etal10,
Bodo_etal11};
\item
linear coupling of different non-axisymmetric modes/branches
\citep[like the transient growth, this coupling is also induced by
the non-normality of the disc shear flow, see e.g.,][]{Bodo_etal05};
\item
viscous and diffusive effects \citep[e.g.,][]{Sano_etal98,
Sano_Stone02, Fleming_Stone03, Fromang_etal07, Pessah_Chan08,
Latter_etal09, Longaretti_Lesur10, Pessah_Goodman09}.
\end{enumerate}
Certainly, consistent numerical simulations are able to cover the
whole set of the above processes, however in simulations all the
processes are intermixed and, in order to understand the dynamics,
one has to find a way to disentangle them. In analytical studies,
one is compelled to restrict attention only to some of the above
processes. For instance, the basic scenario in the above-mentioned
several recent papers focusing basically on I, III and VI processes
is the following: one considers the exponential growth of
axisymmetric perturbations, the fastest growing of which gives rise,
in the non-linear regime, to the so-called channel solution. In
essence, this is  process I. Afterwards, on top of this channel
flow, a secondary instability develops and non-axisymmetric,
so-called parasitic, modes grow. This analysis should mimic process
III, albeit with some simplifications, like neglecting the basic
flow shear and Coriolis force, due to  the complexity of the
secondary instability. Dissipative effects (process VI) have been
also considered, but processes II, IV and V have been in general
neglected, however, there are cases where they may be appreciable
and even decisive. The goal of this paper is in fact the analysis of
the importance of processes II and V, through a study of the linear
dynamics of perturbation in an incompressible stratified disc flow
with a vertical magnetic field. A detailed numerical investigation
of the non-linear process IV is out of the scope of the present
study and will be covered in a subsequent paper.

We would like to draw particular attention to process V -- a linear
coupling of non-axisymmetric modes. This phenomenon, caused by shear
flow non-normality, was revealed in the 1990's and is widely
accepted by the hydrodynamical and geophysical/meteorological
communities \citep[e.g.,][]{Chagelishvili_etal97b,
Farrell_Ioannou00, Vanneste_Yavneh04,
Olafsdotir_etal08,Bakas_Farrell09, Kalashnik_etal06}. It is even
more significant in astrophysical disc flows, where background
differential rotation (i.e., flow non-normality) is strong and,
hence, the linear mode coupling is an inevitable occurrence.
Examples of the studies taking into account the coupling in
astrophysical discs are \cite{Tevzadze_etal03, Bodo_etal05,
Tevzadze_etal08, Mamatsashvili_Chagelishvili07,
Heinemann_Papaloizou09a, Tevzadze_etal10, Paardekooper_etal10,
Mamatsashvili_Rice11}. This paper also demonstrates yet another
manifestation of the coupling phenomenon in magnetized discs.

In the present paper, we primarily perform a comparative analysis of
the exponential instability of axisymmetric perturbations and
transient growth of non-axisymmetric ones, i.e., a quantitative
comparison of the processes I and II. We also focus on the linear
coupling of various non-axisymmetric physical modes of perturbations
existing in the disc flow, i.e., on the process V. As a matter of
fact, in complex/unstable systems, the dynamics of perturbation
modes is driven by various physical phenomena. For instance, in the
considered here magnetized incompressible disc flow, there are two
types of non-axisymmetric modes: 1. magnetic mode that directly taps
into the energy store of disc flow shear and experiences transient
growth and 2. inertia-gravity waves that only weakly directly
exchange energy with the disc flow \citep{Balbus03, Latter_Balbus09,
Ogilvie98}. However, as we demonstrate here, the latter mode is
coupled linearly with the magnetic mode and draws energy from it. As
a result, inertia-gravity waves can also become energetically
intensive and affect dynamical processes, particularly, angular
momentum transport.

The disc model we consider is similar to that of
\cite{Balbus_Hawley92, Goodman_Xu94,Latter_etal09}, i.e., it
represents a local patch of incompressible, vertically stratified
ionized Keplerian disc penetrated by spatially uniform background
magnetic field. We work in terms of the shearing waves (Kelvin
modes) that are a natural basis for the shearing box models of discs
for describing the dynamics of waves
\citep[e.g.,][]{Goldreich_Tremaine78} and vortices
\citep[e.g.,][]{Lominadze_etal88,Chagelishvili_etal03,Johnson_Gammie05}
as well as their linear coupling
\citep[e.g.,][]{Bodo_etal05,Mamatsashvili_Chagelishvili07,Heinemann_Papaloizou09a}.
As mentioned above, the considered system contains two distinct
physical types/modes of perturbations: magnetic mode and
inertia-gravity waves. In turn, each of these modes consists of two
components. The magnetic mode has two counter propagating components
that exhibit growth due to the MRI, but during a limited time
interval. The second, inertia-gravity mode also consists of two
counter propagating waves that can grow due to the coupling with the
magnetic mode. To analyse the dynamics of these four components, we
reduce the linearized ideal magnetohydrodynamic (MHD) equations in
the shearing box to a set of four first order differential equations
for eigen-variables characterizing each component. Then, analyzing
the dynamics using the eigen-variables, we grasp the transient
growth as well as the mode linear coupling processes. We also
contrast the exponential growth of axisymmetric magnetic mode
perturbations to transient growth of non-axisymmetric ones.

The paper is organized as follows. In Section 2, we describe the
mathematical formalism -- linearize basic equations and cast them
into a system of four first order differential equations for mode
eigen-variables. Section 3 is devoted to the dynamics of magnetic
and inertia-gravity modes and to the comparison of axisymmetric and
non-axisymmetric growths of the dominant magnetic mode. Summary and
discussion are given in Section 4.

\section{Physical model and main equations}

To study the dynamics of perturbation modes in a stratified
Keplerian disc threaded by a vertical magnetic field ${\bf
B}_0=B_{0z}{\bf \hat{z}}$, following
\citet{Balbus_Hawley92,Tagger_etal92,Goodman_Xu94,Latter_etal09,
Pessah_Chan12}, we adopt a local shearing box approximation
\citep{Goldreich_Lynden-Bell65}. In the shearing box model, disc
dynamics is studied in a local Cartesian reference frame co-rotating
with the disc's angular velocity at some fiducial radius $r_0$ from
the central star, so that curvature effects due to cylindrical
geometry of the disc are ignored. In this rotating coordinate frame,
the unperturbed differential rotation of the disc manifests itself
as a parallel azimuthal flow with a constant velocity shear in the
radial direction. The Coriolis force is included to take into
account the effects of the coordinate frame rotation. The vertical
component of the gravity force due to the central star is also
present and responsible for disc's vertical stratification. The
background magnetic field ${\bf B}_0$ is assumed weak (sub-thermal)
in the sense that the Alfv\'{e}n speed is much less than the sound
speed, so that we can work in the Boussinesq approximation, as in
\citet{Balbus_Hawley91,Balbus_Hawley92}. As a result, the basic
dynamical equations of incompressible ideal MHD in the shearing box
are \citep[e.g.,][]{Goodman_Xu94}:
\begin{multline}
\frac{\partial {\bf u}}{\partial t}+({\bf u}\cdot\nabla){\bf
u}+2\Omega{\bf\hat{z}}\times {\bf u}= -\frac{1}{\rho}\nabla
P+\frac{({\bf B}\cdot \nabla){\bf B}}{4\pi\rho}+\\+2q\Omega^2x{\bf
\hat{x}}-\Omega^2z{\bf \hat{z}},
\end{multline}
\begin{equation}
\frac{\partial {\bf B}}{\partial t}=\nabla \times ({\bf u}\times
{\bf B}),
\end{equation}
\begin{equation}
\nabla\cdot{\bf u}=0,
\end{equation}
\begin{equation}
\nabla\cdot{\bf B}=0,
\end{equation}
where ${\bf u}$ is the velocity in the local frame, $\rho$ is the
density, ${\bf B}$ is the magnetic field, $P=p+{\bf B}^2/8\pi$ is
the total pressure, $\Omega$ is the angular velocity of the local
rotating reference frame, equal to the disc's angular velocity at
$r_0$: $\Omega(r_0)$. $x,y,z$ are, respectively, the radial,
azimuthal and vertical coordinates with the corresponding unit
vectors ${\bf \hat{x}}, {\bf \hat{y}}, {\bf \hat{z}}$ in these
directions. The shear parameter $q=1.5$ for the Keplerian
differential rotation considered in this paper.

The set of equations (1)-(4) has an equilibrium solution that is
stationary and axisymmetric. In this unperturbed state, the velocity
field of Keplerian rotation represents, as noted above, a parallel
azimuthal flow, ${\bf u_0}$, with a linear shear in the radial
direction
\[
u_{0x}=u_{0z}=0,~~~u_{0y}=-q\Omega x.
\]
In the shearing box model, equilibrium density $\rho_0$ and total
pressure $P_0$ depend only on the vertical $z-$coordinate and
satisfy the hydrostatic balance equation
\[
\frac{1}{\rho_0}\frac{dP_0}{dz}+\Omega^2z=0,
\]
while the background magnetic field along the $z$-axis, ${\bf
B}_0=B_{0z}{\bf \hat{z}}$, is assumed to be spatially uniform, as is
typical for configurations with non-zero net flux.

\subsection{Perturbation equations}

Consider now perturbations ${\bf u}={\bf u}_0+{\bf u'}$,
$\rho=\rho_0+\rho'$, $P=P_0+P'$, ${\bf B}={\bf B}_0+{\bf B'}$ to the
equilibrium described above. Assuming the perturbation of density to
be small, $\rho'\ll \rho_0$, from equations (1)-(4) we obtain the
following system governing the perturbation dynamics
\begin{multline}
\frac{D{\bf u'}}{Dt}+({\bf u'}\cdot\nabla){\bf
u'}+2\Omega{\bf\hat{z}}\times {\bf u'}-q\Omega u'_x{\bf
\hat{y}}=-\frac{1}{\rho_0}\nabla
P'+\\+\frac{\rho'}{\rho_0^2}\frac{dP_0}{dz}{\bf
\hat{z}}+\frac{B_{0z}}{4\pi\rho_0}\frac{\partial {\bf B'}}{\partial
z}+\frac{({\bf B'}\cdot \nabla){\bf B'}}{4\pi\rho_0},
\end{multline}
\begin{equation}
\frac{D{\bf B'}}{Dt}=B_{0z}\frac{\partial {\bf u'}}{\partial
z}-q\Omega B'_x{\bf \hat{y}}+({\bf B'}\cdot\nabla){\bf u'}-({\bf
u'}\cdot\nabla){\bf B'}
\end{equation}
\begin{equation}
\nabla\cdot{\bf u'}=0,
\end{equation}
\begin{equation}
\nabla\cdot{\bf B'}=0,
\end{equation}
where
\[
\frac{D}{Dt} \equiv \frac{\partial}{\partial t}-q\Omega x
\frac{\partial}{\partial y}.
\]
The form of equations (5)-(8) permits decomposition of the
perturbations into shearing plane waves, or spatial Fourier
harmonics (SFHs) with time-dependent amplitudes and phases
\begin{equation}
F({\bf r},t)=\bar{F}(t){\rm exp}[{\rm i}k_x(t)x+{\rm i}k_yy+{\rm
i}k_zz],
\end{equation}
\[
k_x(t)=k_x(0)+q\Omega k_y t,
\]
where $F\equiv ({\bf u}',\rho',P',{\bf B}')$ denotes the perturbed
quantities and $\bar{F}\equiv(\bar{\bf
u},\bar{\rho},\bar{P},\bar{\bf B})$ are the amplitudes of
corresponding SFHs. The azimuthal, $k_y$, and vertical, $k_z$,
wavenumbers remain unchanged, whereas the radial wavenumber $k_x(t)$
varies with time at a constant rate $q\Omega k_y$ if $k_y\neq 0$
(i.e., for non-axisymmetric perturbations) due to sweeping of wave
crests by the background shear flow. In other words, an initially
leading SFH (with $k_x(0)/k_y < 0$) eventually becomes trailing
(with $k_x(t)/k_y > 0$) as time goes by. This change of SFH's
orientation from leading to trailing at $k_x(t)=0$ is called
`swing'. In the local disc model, shearing plane waves are a natural
platform of decomposition of perturbations compatible with the
shearing sheet boundary conditions \citep[see
e.g.,][]{Hawley_etal95,Heinemann_Papaloizou09b} and, in fact,
represent the basic/simplest 'elements' of dynamical processes at
linear shear \citep{Yoshida05}.

Substituting (9) into equations (5)-(8), all the non-linear terms
vanish identically in these equations and we arrive at the following
system of linear first order ordinary differential equations that
govern the dynamics of SFHs of perturbations
\begin{equation}
\frac{d\bar{u}_x}{dt}=-\frac{{\rm i}k_x(t)}{\rho_0}\bar{P}+2\Omega
\bar{u}_y+\frac{{\rm i}k_zB_{0z}}{4\pi\rho_0}\bar{B}_x,
\end{equation}
\begin{equation}
\frac{d\bar{u}_y}{dt}=-\frac{{\rm i}k_y}{\rho_0}\bar{P}+(q-2)\Omega
\bar{u}_x+\frac{{\rm i}k_zB_{0z}}{4\pi\rho_0}\bar{B}_y,
\end{equation}
\begin{equation}
\frac{d\bar{u}_z}{dt}=-\frac{{\rm
i}k_z}{\rho_0}\bar{P}+\frac{\bar{\rho}}{\rho_0^2}\frac{dP_0}{dz}+\frac{{\rm
i}k_zB_{0z}}{4\pi\rho_0}\bar{B}_z,
\end{equation}
\begin{equation}
\frac{d\bar{B}_x}{dt}={\rm i}k_zB_{0z}\bar{u}_x,
\end{equation}
\begin{equation}
\frac{d\bar{B}_y}{dt}={\rm i}k_zB_{0z}\bar{u}_y-q\Omega \bar{B}_x,
\end{equation}
\begin{equation}
k_x(t)\bar{u}_x+k_y\bar{u}_y+k_z\bar{u}_z=0,
\end{equation}
\begin{equation}
k_x(t)\bar{B}_x+k_y\bar{B}_y+k_z\bar{B}_z=0.
\end{equation}
We note that vanishing of non-linear terms for perturbations of a
single Fourier harmonic form (9), that reduces full non-linear
equations exactly to linear ones, is a general property of
incompressible MHD discovered by \citet{Goodman_Xu94}. Thus,
equations (10)-(16) and the subsequent analysis are, in fact, valid
for any values of the amplitudes $\bar{{\bf u}}$ and $\bar{{\bf
B}}$.

Using the continuity equation, the density amplitude $\bar{\rho}$
can be found through the $z$-velocity amplitude
\[
\frac{d\bar{\rho}}{dt}=-\bar{u}_z\frac{d\rho_0}{dz}
\]
and then from equations (13)-(16), be expressed through the vertical
magnetic field,
\begin{equation}
\bar{\rho}=\frac{{\rm i}\bar{B}_z}{B_{0z}k_z}\frac{d\rho_0}{dz}.
\end{equation}
Before proceeding further, we make a non-dimensionalization of
quantities in terms of the orbital time $\Omega^{-1}$, sound speed
$c_s$ and scale-height $H=c_s/\Omega$ specific to the disc's local
dynamics:
\[
\Omega t\rightarrow t,~~~\frac{z}{H}\rightarrow z,~~~
\frac{P_0}{\rho_0 c_s^2}\rightarrow P_0,~~~H{\bf k}\rightarrow {\bf
k},
\]
\[
\frac{\bar{\rho}}{\rho_0}\rightarrow
\bar{\rho},~~~\frac{\bar{P}}{\rho_0 c_s^2}\rightarrow \bar{P},
~~~\frac{\bar{{\bf u}}}{c_s}\rightarrow \bar{{\bf
u}},~~~\frac{\bar{{\bf B}}}{B_{0z}}\rightarrow \bar{\bf B}.
\]
Despite working in the incompressible limit, we still prefer to use
the sound speed instead of the Alfv\'{e}n speed $v_{Az}=B_{0z}/(4\pi
\rho_0)^{1/2}$ in the normalizations. A more natural normalization
length scale to be used in the present incompressible problem would
obviously be a characteristic length $H_m=v_{Az}/\Omega$ along the
$z-$axis associated with the background magnetic field, but the
reason for using the scale-height, $H$, instead is that this will be
useful below in making a connection with the compressible case by
means of the plasma $\beta=8\pi \rho_0 c_s^2/B_{0z}^2=2H^2/H_m^2$
parameter. Typically, this parameter is large for weak fields (i.e.,
$H_m\ll H$) and we take a value $\beta=400$ often adopted in
simulations of MRI-driven turbulence in discs
\citep[e.g.,][]{Fromang_Papaloizou07,Guan_etal09}.

In the Boussinesq approximation employed here, all the equilibrium
quantities and their vertical derivatives entering equations
(10)-(17) are approximately uniform in the vertical $z-$coordinate.
This holds if the vertical length-scale of perturbations is much
shorter than the disc scale-height, implying that our incompressible
analysis is, strictly speaking, self-consistent for $k_z\gg 1$ with
the above scaling. Still, as will be shown below, it can give an
idea about the trend of mode dynamics (amplification) at $k_z \gsim
1$, near the compressibility regime.

Next, substituting the velocities from equations (13)-(15) and
density from equation (17) into equations (10)-(12), we arrive at
the set of second order differential equations only for the magnetic
field components, which after the above non-dimensionalizations take
the form (henceforth bars over the amplitudes will be omitted)
\begin{equation}
\frac{d^2B_x}{dt^2}=2\frac{dB_y}{dt}+k_zk_x(t)P+\left(2q-\frac{2k_z^2}{\beta}\right)B_x,
\end{equation}
\begin{equation}
\frac{d^2B_y}{dt^2}=-2\frac{dB_x}{dt}+k_zk_yP-\frac{2k_z^2}{\beta}B_y,
\end{equation}
\begin{equation}
\frac{d^2B_z}{dt^2}=k_z^2P-\left(N^2+\frac{2k_z^2}{\beta}\right)B_z,
\end{equation}
supplemented with the divergence-free constraint
\begin{equation}
k_x(t)B_x+k_yB_y+k_zB_z=0,
\end{equation}
where $N^2=-d~{\rm ln}~\rho_0/d~{\rm ln}~z$ is the squared
Brunt-V\"{a}is\"{a}l\"{a} frequency normalized by $\Omega^2$; it is
positive here and therefore the disc is convectively stable.
Everywhere below we mostly use a value $N^2=0.8$, although show that
the results do not change much with different stratifications.
Equations (18)-(21) form the basis for our subsequent analysis. They
contain all the information on various perturbation modes existing
in the shearing box model of a stratified, incompressible disc
threaded by a constant background vertical magnetic field. We will
classify these modes below by making use of the WKBJ approach.

\subsection{Total perturbation energy and $\alpha$ parameter}

For further analysis it is useful to introduce the total spectral
energy density of SFHs and derive its evolution equation. From
equations (10)-(17), we obtain the following equation governing the
energy evolution \citep[see e.g.,][]{Balbus03}
\begin{equation}
\frac{dE}{dt}=q\left[u_xu_y^{\ast}+u_x^{\ast}u_y-\frac{2}{\beta}\left(B_xB_y^{\ast}+B_x^{\ast}B_y\right)\right]=q\alpha,
\end{equation}
where asterisks denote complex conjugate and the non-dimensional
total spectral energy density $E=E_k+E_g+E_m$ is composed of three
basic terms: the kinetic energy density
\[
E_k=|u_x|^2+|u_y|^2+|u_z|^2,
\]
the potential, or buoyancy energy density
\[
E_g=\rho_0\left(\frac{dP_0}{dz}/\frac{d\rho_0}{dz}\right)|\rho|^2=\frac{N^2}{k_z^2}|B_z|^2
\]
and the magnetic energy density
\[
E_m=\frac{2}{\beta}(|B_x|^2+|B_y|^2+|B_z|^2).
\]
The right hand side of equation (22) is the shear parameter
multiplied by the sum of Reynolds and Maxwell stresses, that is, by
the well-known Shakura-Sunyaev $\alpha$ parameter. These stresses
ensure energy exchange process between the disc mean flow and
perturbations; in the absence of the shear (i.e., in a rigidly
rotating disc), the total energy of perturbations is conserved. So,
the growth of small perturbations, and therefore the MRI, is
entirely fed with the energy drawn from the disc's differential
rotation by means of the stresses.

\subsection{Eigen-variables and classification of modes -- WKBJ approximation}

The shear parameter $q$ enters equations (18)-(21) implicitly
through the time-dependent radial wavenumber $k_x(t)$ and explicitly
in the right hand side of equation (18). Therefore, we can
distinguish between two kinds of effects induced by the disc flow
shear on the perturbation dynamics. The implicit dependence on the
shear leads to finite-time phenomena -- transient growth and
coupling of different modes  -- in the dynamics of non-axisymmetric
perturbations that arise from the drift (time-variation) of $k_x(t)$
in the wavenumber space (see next subsection); they vanish for
axisymmetric perturbations with $k_y=0$. The explicit dependence on
the shear is not related to non-axisymmetry and equally affects both
axisymmetric and non-axisymmetric perturbations. This term sets the
frequency of epicyclic oscillations and, in particular, is
responsible for the existence of the MRI in differentially rotating
discs.

To classify and characterize all the modes involved in equations
(18)-(21), first we consider the regime where the radial wavenumber
is much larger than the azimuthal one, $|k_y| \ll |k_x(t)|$. This
condition implies that the time-dependent radial wavenumber $k_x(t)$
varies a little during the shear/dynamical time,
$(q\Omega)^{-1}|dk_x/dt|=|k_y|\ll|k_x(t)|$, so that the WKBJ, or
adiabatic approximation holds with respect to the small parameter
$k_y/k_x$ and hence the effect of disc flow shear stemming from
non-axisymmetry of perturbations does not play a role. We will see
later that in the opposite regime $|k_y|\gsim |k_x(t)|$ (which
ensues eventually as $k_x(t)$ drifts along the $k_x-$axis), the
dynamics is non-adiabatic (non-WKBJ) and shear-induced transient
effects become important. So, in the present case, we ignore for the
moment the variation of $k_x$ with time in equations (18)-(21),
retain only the explicit shear term and look for solutions in the
standard form $\propto {\rm exp}\left(-{\rm i}\int^t\omega(t')
dt'\right)$. Substituting this into equations (18)-(21) and assuming
that the time-dependent frequency, $\omega(t)$, also satisfies the
same WKBJ condition $|d\omega(t)/dt|\ll\omega^2(t)$ for $|k_y|\ll
|k_x|$, we get a linear system of equations
\begin{equation}
\left(\omega^2+2q-\frac{2k_z^2}{\beta}\right)B_x-2{\rm i}\omega
B_y+k_zk_xP=0
\end{equation}
\begin{equation}
2{\rm i}\omega
B_x+\left(\omega^2-\frac{2k_z^2}{\beta}\right)B_y+k_zk_yP=0
\end{equation}
\begin{equation}
\left(\omega^2-N^2-\frac{2k_z^2}{\beta}\right)B_z+k_z^2P=0
\end{equation}
\begin{equation}
k_xB_x+k_yB_y+k_zB_z=0.
\end{equation}
Associated eigenfrequencies are obtained by equating the determinant
to zero,
\begin{equation*}
\left|\begin{matrix}
  \omega^2+2q-\frac{2k_z^2}{\beta} & -2{\rm i}\omega & 0 & k_xk_z \\
  2{\rm i}\omega & \omega^2-\frac{2k_z^2}{\beta} & 0 & k_yk_z\\
  0 & 0 & \omega^2-N^2-\frac{2k_z^2}{\beta} & k_z^2\\
  k_x & k_y & k_z & 0
\end{matrix}\right|=0,
\end{equation*}
which yields the formal WKBJ dispersion relation:
\begin{equation}
\omega^4-a_1\omega^2-a_2=0,
\end{equation}
\[
a_1=N^2\frac{k_y^2+k_x^2}{k^2}+2(2-q)\frac{k_z^2}{k^2}+\frac{4k_z^2}{\beta}-
2q\frac{k_y^2}{k^2},
\]
\[
a_2=2qN^2\frac{k_y^2}{k^2}-\frac{4k_z^4}{\beta^2}-\frac{2k_z^2}{\beta}\left(
N^2\frac{k_x^2+k_y^2}{k^2}-2q\frac{k_y^2+k_z^2}{k^2}\right)
\]
with $k^2=k_x^2+k_y^2+k_z^2$. In the limit $k_y=0$, equation (27)
reduces to the well-known dispersion relation of
\citet{Balbus_Hawley91} describing the axisymmetric MRI. So, we can
view it as a generalized non-axisymmetric WKBJ dispersion relation
at small $k_y$.

Dispersion relation (27) is a quadratic equation with respect to
$\omega^2$ and has two distinct solutions corresponding to two
physically different types of perturbation modes:
\begin{enumerate}[1.]
\item
An inertia-gravity wave with the frequency
\begin{equation}
\omega_g^2=\frac{a_1}{2}+\sqrt{a_2+\frac{a_1^2}{4}},
\end{equation}
the restoring force for which is mainly provided by rotation and
vertical gravity (buoyancy), but is modified by magnetic field
tension. For very large $|k_x|\gg|k_y|,|k_z|$ the effect of rotation
is negligible and this frequency asymptotically tends to
$\omega_g^2\approx N^2+2k_z^2/\beta$. Generally, inertia-gravity
waves play an important role in overall disc dynamics and they were
studied extensively in the past mainly in the context of
unmagnetized discs \citep{Lubow_Pringle93, Korycansky_Pringle95,
Ogilvie98, Balbus03, Tevzadze_etal03,Tevzadze_etal08,
Latter_Balbus09}. Obviously, these waves have $\omega_g^2>0$, since
the disc is stratified stably.
\item
A magnetic mode with the frequency/growth rate\footnote{In the
literature, this mode is often called MRI mode, but we prefer to
call it magnetic mode \citep[see also][]{Ogilvie98}, because apart
from the MRI at smaller radial wavenumbers, it also displays stable
oscillations with Alfv\'{e}n frequency at larger radial
wavenumbers.}
\begin{equation}
\omega_m^2=\frac{a_1}{2}-\sqrt{a_2+\frac{a_1^2}{4}},
\end{equation}
the restoring force for which is mainly provided by magnetic
tension, but modified by rotation and vertical buoyancy. Due to the
disc's differential rotation, this mode becomes unstable
($\omega_m^2<0$), that is, exhibits the MRI when $k_z^2 < q\beta$,
at radial wavenumbers smaller than a certain critical value,
\begin{equation}
k_x^2<\left(q\beta-k_z^2\right)\frac{\beta N^2
k_y^2+2k_z^2(k_y^2+k_z^2)}{k_z^2(\beta N^2+2k_z^2)},
\end{equation}
as it follows from expression (29). At larger $k_x$, the magnetic
mode is stable and oscillatory; it is essentially of the same nature
as Alfv\'{e}n waves in classical MHD. Again, in the limit of very
large radial wavenumbers $|k_x|\gg|k_y|,|k_z|$ the effect of
rotation is negligible in the dynamics of the magnetic mode and its
oscillation frequency tends to Alfv\'{e}nic, $\omega_m^2\approx
2k_z^2/\beta$.
\end{enumerate}
Thus, in a stratified incompressible magnetized disc, perturbations
can be classified into two basic types -- the inertia-gravity wave
and magnetic modes and any general perturbation can be decomposed
into the sum of these two modes.

Eigenvalue equations (23)-(26) can be transformed into a slightly
different form. We eliminate the total pressure $P$ through $B_x$
and $B_y$ from equations (25)-(26) and switch to new independent
variables
\[h_1=B_x,~~~h_2=dB_x/dt,~~~h_3=B_y,~~~h_4=dB_y/dt,
\]
which will be our basic variables in the subsequent analysis. After
substituting these into equations (23)-(24), we obtain
\begin{equation}
\frac{d{\bf h}}{dt}=-{\rm i}\omega{\bf h}={\bf A}\cdot{\bf h},
\end{equation}
where ${\bf h}=[h_1, h_2, h_3, h_4]^T$ is the state column vector
and the components of a new matrix ${\bf A}$ are
\begin{align*}
A_{11}&=A_{13}=A_{14}=0,~~~~~A_{12}=1, \\
A_{21}&=2q-\frac{2k_z^2}{\beta}-(N^2+2q)\frac{k_x^2}{k^2},~~~~~~A_{22}=2\frac{k_xk_y}{k^2},\\
A_{23}&=-N^2\frac{k_xk_y}{k^2},~~~~~~~~~~~~~A_{24}=2\left(1-\frac{k_x^2}{k^2}\right),\\
A_{31}&=A_{32}=A_{33}=0,~~~~~A_{34}=1\\
A_{41}&=-(N^2+2q)\frac{k_xk_y}{k^2},~~~~~~~~A_{42}=-2\left(1-\frac{k_y^2}{k^2}
\right),\\
A_{43}&=-\frac{2k_z^2}{\beta}-N^2\frac{k_y^2}{k^2},~~~~~~~~~~~~A_{44}=-2\frac{k_xk_y}{k^2}.
\end{align*}
Eigenfrequencies associated with equation (31) are found from
\[
{\rm Det}({\bf A}+{\rm i}\omega{\bf I})=0.
\]
Calculating the determinant obviously gives the same dispersion
relation (27). So, the eigenfrequencies $\pm \omega_m$ and $\pm
\omega_g$, given by (28) and (29), multiplied by $-{\rm i}$
represent the eigenvalues of the matrix ${\bf A}$. Since the modes'
frequencies and therefore the eigenvalues are generally distinct and
corresponding eigenvectors are linearly independent, we can use the
spectral decomposition, or diagonalization of ${\bf A}$
\citep{Golub_VanLoan96},
\begin{equation}
{\bf A}={\bf C}{\bf \Lambda}{\bf C^{-1}},
\end{equation}
where ${\bf \Lambda}$ is the diagonal matrix formed from the
eigenvalues ${\rm i}\omega_m, -{\rm i}\omega_m, {\rm i}\omega_g,
-{\rm i}\omega_g$ of the ${\bf  A}$ matrix,
\[
{\bf \Lambda}=\begin{pmatrix} {\rm i}\omega_m & 0 & 0 & 0 \\
  0 & -{\rm i}\omega_m & 0 & \\
  0 & 0 & {\rm i}\omega_g & 0\\
  0 & 0 & 0 & -{\rm i}\omega_g
\end{pmatrix}
\]
and the columns of a matrix ${\bf C}$ are the eigenvectors of ${\bf
A}$, such that the first column corresponds to the eigenvalue ${\rm
i}\omega_m$, second column to $-{\rm i}\omega_m$, third to ${\rm
i}\omega_g$ and fourth to $-{\rm i}\omega_g$,
\[
{\bf  C}=\begin{pmatrix} 1 & 1 & 1 & 1\\ \\
{\rm i}\omega_m & -{\rm i}\omega_m & {\rm i}\omega_g & -{\rm
i}\omega_g\\ \\
G_1(\omega_m) & G_1(-\omega_m) & G_1(\omega_g) & G_1(-\omega_g)\\ \\
\\ G_2(\omega_m) & G_2(-\omega_m) & G_2(\omega_g) & G_2(-\omega_g)
\end{pmatrix}
\]
where the auxiliary functions $G_1(\omega)$ and $G_2(\omega)$ are
defined as
\[
G_1(\omega) \equiv-\frac{A_{41}+{\rm i}\omega
A_{42}}{\omega^2+A_{43}+{\rm i}\omega A_{44}},~~~~~G_2(\omega)
\equiv i\omega G_1(\omega).
\]
An advantage of the spectral decomposition is that it allows us to
reduce equation (31) to a set of decoupled first order differential
equations, i.e., to corresponding canonical equations. Substituting
(32) into equation (31) and defining a new vector variable
${\boldsymbol \psi}\equiv[\psi^{(+)}_m, \psi^{(-)}_m, \psi^{(+)}_g,
\psi^{(-)}_g]^T$ as ${\boldsymbol \psi}={\bf C}^{-1}\cdot{\bf h}$,
we obtain the canonical form of equation (31) for ${\boldsymbol
\psi}$,
\begin{equation}
\frac{d{\boldsymbol \psi}}{dt}={\bf \Lambda}\cdot {\boldsymbol
\psi}.
\end{equation}
Because ${\bf \Lambda}$ is diagonal, the functions $\psi^{(+)}_m,
\psi^{(-)}_m, \psi^{(+)}_g, \psi^{(-)}_g$ are dynamically decoupled
and evolve independently from each other, respectively, with the
frequencies $\omega_m, -\omega_m, \omega_g, -\omega_g$. Physically,
they characterize the above-classified modes with corresponding
counter-propagating (denoted by '+' and '-') components --
$\psi^{(+)}_m$, $\psi^{(-)}_m$ describe the magnetic mode and
$\psi^{(+)}_g$, $\psi^{(-)}_g$ describe inertia-gravity waves -- and
in this respect are very convenient and revealing. Below we will
refer to them as mode eigen-variables. The absence of coupling among
modal equations (33) in the linear analysis means that in the WKBJ
limit, the perturbation modes evolve without exchanging energy among
each other, i.e., initially exciting one either mode with a specific
characteristic time-scale does not lead to the excitation of other
modes with different time-scales. In the following subsection, we
will see that in the non-WKBJ regime, these equations are no longer
independent from each other, i.e. become coupled that, in turn,
results in the mutual linear coupling between inertia-gravity waves
and the magnetic mode.

\subsection{The dynamical equations in the non-WKBJ regime}

In the WKBJ/adiabatic regime considered above, the radial wavenumber
$k_x(t)$ and hence the mode eigenfrequencies have been assumed
approximately independent of time by virtue of the limit $|k_y|\ll
|k_x(t)|$. However, after some time, $k_x(t)$, drifting along the
$k_x-$axis, enters the range $|k_y|\gsim |k_x|$, where the WKBJ
treatment breaks down and one must take into account, neglected so
far, the dependence of $k_x(t)$ on time due to the disc flow shear
in equations (18)-(21). In other words, in this
non-WKBJ/non-adiabatic regime, transient effects arising from the
implicit presence of the shear in equations (18)-(21) manifest
themselves in the mode dynamics.

\begin{figure}
\includegraphics[width=\columnwidth]{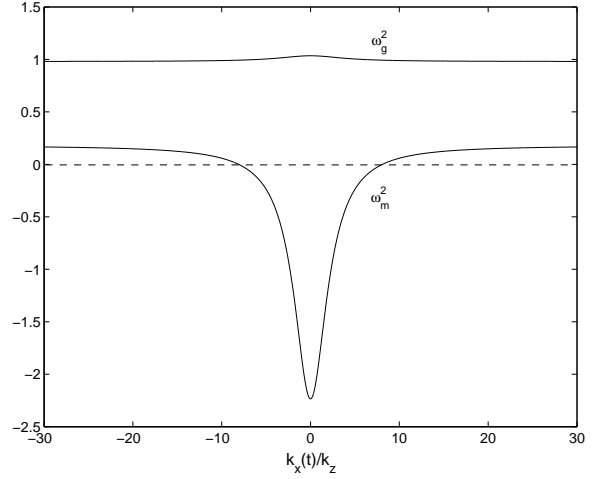} \caption{Squared
frequencies $\omega^2_m$ and $\omega^2_g$ from equations (28) and
(29), respectively, versus the time-dependent ratio $k_x(t)/k_z$ at
$k_y/k_z=2, k_z=6$ and $N^2=0.8$. $\omega^2_m$ is negative during a
finite time, in the interval $|k_x(t)/k_z|<8.1$ (i.e.,
$|k_x(t)/k_y|<4.05$ as given by condition 30), indicating the MRI of
the non-axisymmetric magnetic mode. At large $|k_x(t)|\rightarrow
\infty$, both frequencies go to constant values, $\omega^2_m \approx
2k_z^2/\beta=0.18$ and $\omega^2_g \approx N^2+2k_z^2/\beta=0.98$.}
\end{figure}
\begin{figure}
\includegraphics[width=1.1\columnwidth]{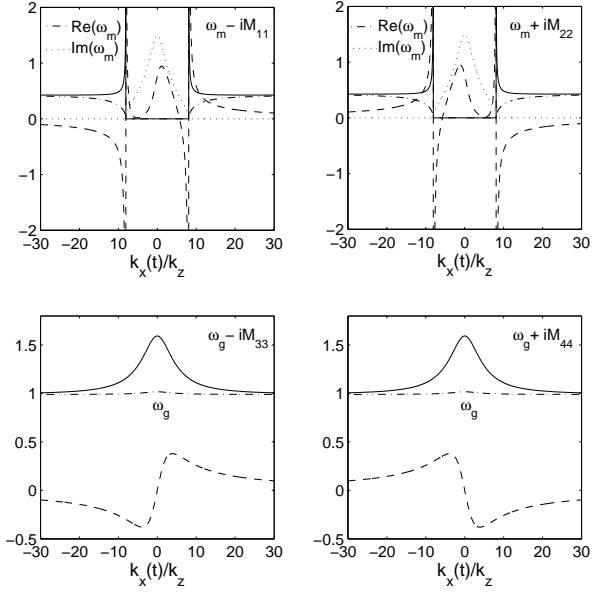}
\caption{Real (solid lines) and imaginary (dashed lines) parts of
the effective, or modified frequencies/growth rates, $\omega_m-{\rm
i}M_{11},~\omega_m+{\rm i}M_{22},~\omega_g-{\rm
i}M_{33},~\omega_g+{\rm i}M_{44}$, versus $k_x(t)/k_z$ at the same
$k_y/k_z, k_z$ and $N^2$ as in Fig. 1. For comparison, we also plot
${\rm Re}(\omega_m)$ (dashed-dotted line) and ${\rm Im}(\omega_m)$
(dotted line) in the upper two panels and purely real $\omega_g$
(dashed-dotted line) in the lower two panels. Sharp peaks near
$k_x/k_z={\pm 8.1} (k_x/k_y={\pm} 4.05)$ in the upper two panels
correspond to the points $\omega_m^2=0$, where the WKBJ
approximation breaks down and the mode dynamics becomes
non-adiabatic. Consequently, the difference between the effective
frequencies/growth rates and $\omega_m, \omega_g$ found in the WKBJ
limit is most appreciable in the non-adiabatic region, between these
two peaks.}
\end{figure}

Combining equations (18)-(21), we eliminate the total pressure
\begin{multline*}
P=-\frac{1}{k_zk^2(t)}\left[2(q-1)k_y\frac{dB_x}{dt}+2k_x(t)\frac{dB_y}{dt}+\right.\\
\left.+(N^2+2q)k_x(t)B_x+N^2k_yB_y\right],
\end{multline*}
where $k^2(t)=k_x^2(t)+k_y^2+k_z^2$. Substituting this, together
with $B_z$ from equation (21), back into equations (18) and (19), we
finally arrive at the system of two second order differential
equations
\begin{multline}
\frac{d^2B_x}{dt^2}=-\frac{2(q-1)
k_yk_x(t)}{k^2(t)}\frac{dB_x}{dt}+2\left[1-\frac{k_x^2(t)}{k^2(t)}\right]\frac{dB_y}{dt}+\\
+\left[2q-\frac{2k_z^2}{\beta}-(N^2+2q)\frac{k_x^2(t)}{k^2(t)}\right]B_x-N^2\frac{k_yk_x(t)}{k^2(t)}B_y,
\end{multline}
\begin{multline}
\frac{d^2B_y}{dt^2}=-2\left[1+\frac{(q-1)k_y^2}{k^2(t)}\right]\frac{dB_x}{dt}-\frac{2
k_yk_x(t)}{k^2(t)}\frac{dB_y}{dt}-\\-(N^2+2q)\frac{k_yk_x(t)}{k^2(t)}B_x-\left[\frac{2k_z^2}{\beta}+N^2\frac{k_y^2}{k^2(t)}\right]B_y.
\end{multline}
Equations (34) and (35) are essentially the same as the central
working equations (2.19) and (2.20) of \citet{Balbus_Hawley92}, but
cast instead in terms of the components $B_x$ and $B_y$. They permit
to more easily and conveniently carry out parallel analysis of
non-axisymmetric and axisymmetric dynamics (in the latter case there
is no need for an involved procedure for recovering the axisymmetric
limit as done in \citet{Balbus_Hawley92}). In a vertically
stratified incompressible disc, there is no characteristic
horizontal length-scale and therefore $k_x$ and $k_y$ enter
equations (34) and (35) in combination $k_x/k$ and $k_y/k$, whereas
$k_z$ enters explicitly as multiplied by $H_m$. So, these equations
in fact contain three free parameters, $k_x/k_z, k_y/k_z$ and $k_z$
by means of which we will describe the dynamics of the modes.
Without loss of generality, we can take $k_y$ and $k_z$ to be
positive, $k_y,k_z>0$.

Equations (34) and (35) rewritten in terms of the variables $h_1,
h_2, h_3, h_4$ introduced above take the form
\begin{equation}
\frac{d{\bf h}}{dt}=[{\bf A}(t)+{\bf A}_1(t)]\cdot{\bf h},
\end{equation}
where ${\bf A}$ is the former matrix as in equation (31) except now
time-dependent through $k_x(t)$. ${\bf A}_1$ is a new matrix
originating from the time-dependence of $k_x$ and is therefore
proportional to the shear parameter $q$,
\[
{\bf A}_1=-\frac{2q}{k^2(t)}\begin{pmatrix} 0 & 0 & 0 & 0\\
0 & k_yk_x(t) & 0 & 0\\
0 & 0 & 0 & 0\\
0 & k_y^2 & 0 & 0.
\end{pmatrix}
\]
We see that in the non-WKBJ regime, where the variation of $k_x$
with time is allowed for, new shear-related terms appear in basic
equation (36) compared to its counterpart WKBJ equation (31); these
terms are proportional to $k_y/k$ and vanish in the WKBJ limit
and/or for axisymmetric perturbations.

\begin{figure*}
\centering\includegraphics[width=0.97\textwidth,height=0.7\textwidth]{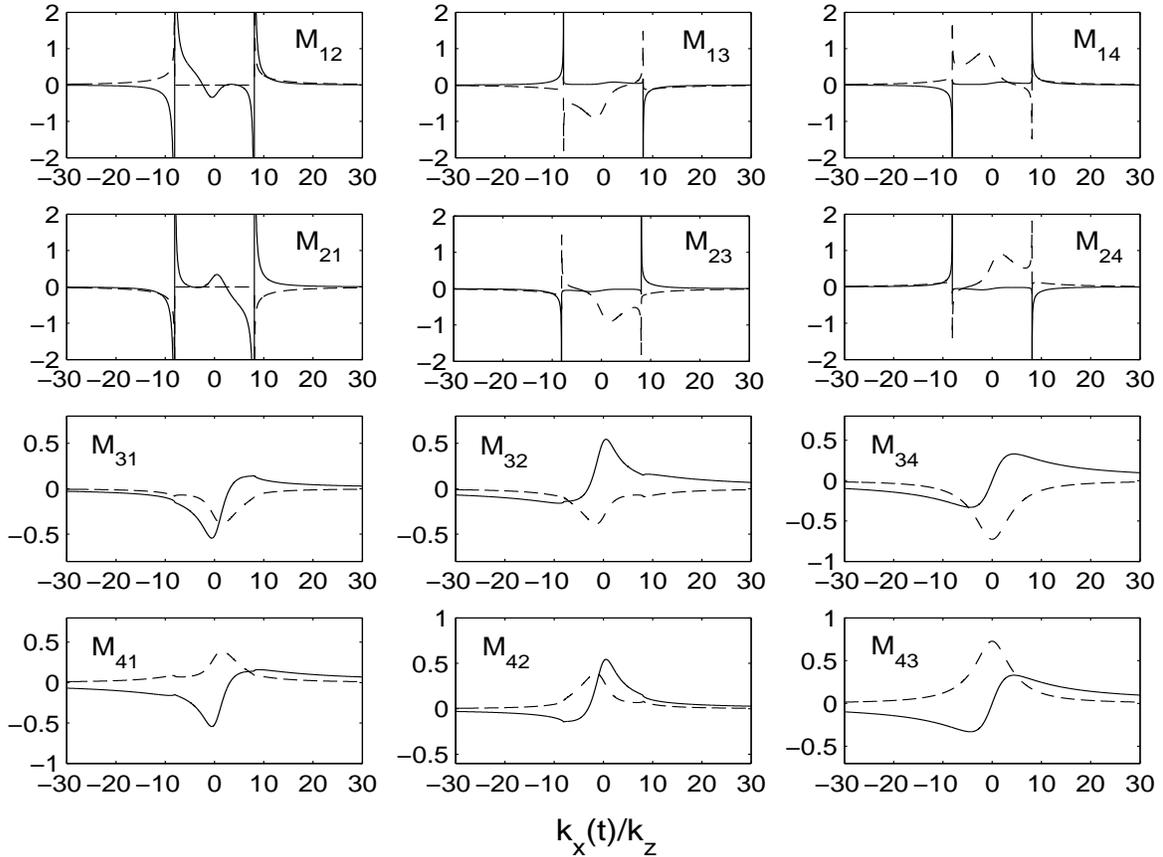}
\caption{The non-diagonal elements of the coupling matrix ${\rm M}$
versus $k_x(t)/k_z$ at the same $k_y/k_z, k_z$ and $N^2$ as in Fig.
1. In each panel, solid line shows real and dashed line imaginary
parts of the corresponding matrix element. These coupling
coefficients vanish at large $|k_x(t)/k_z|\gg 1$, in the WKBJ regime
where the dynamics is adiabatic, but are appreciable, and thus
non-adiabatic transient effects due to the disc flow shear are most
effective, at intermediate and small radial wavenumbers,
$|k_x(t)/k_z| \lsim 10$.}
\end{figure*}

To extend canonical equations (33) to the non-WKBJ case, we should
take into account that the matrix ${\bf C}$ in the spectral
decomposition of ${\bf A}(t)$ also varies with time through
$k_x(t)$. As a consequence, substituting representation (32) into
equation (36), we obtain a more general set of canonical equations
governing the linear dynamics of eigen-variables in both adiabatic
and non-adiabatic regimes
\begin{equation}
\frac{d{\boldsymbol \psi}}{dt}=[{\bf \Lambda}(t)+{\bf
M}(t)]\cdot{\boldsymbol \psi},
\end{equation}
where the generalized eigen-variables are given, as before, by
${\boldsymbol \psi}(t)={\bf C}^{-1}(t)\cdot{\bf h}(t)$. They are a
natural extension of the eigen-variables defined above in the WKBJ
regime. ${\bf \Lambda}(t)$ is the former eigenvalue matrix which has
also become time-dependent through $k_x(t)$. The matrix
\[
{\bf M}(t)={\bf C}^{-1}(t){\bf A}_1(t){\bf C}(t)-{\bf
C}^{-1}(t)\frac{d}{dt}{\bf C}(t)
\]
is a new term, compared to canonical WKBJ equations (33), introduced
by the shear. Like ${\bf A}_1$, it is also proportional to $k_y/k$
and thus was absent in the WKBJ regime considered above. In order to
better understand its role in the mode dynamics, it is convenient to
write equation (37) componentwise
\begin{equation*}
\frac{d\psi^{(+)}_m}{dt}-{\rm i}(\omega_m-{\rm i}M_{11})\psi^{(+)}_m
=M_{12}\psi^{(-)}_m+M_{13}\psi^{(+)}_g+M_{14}\psi^{(-)}_g,
\end{equation*}
\begin{equation}
\frac{d\psi^{(-)}_m}{dt}+{\rm i}(\omega_m+{\rm i}M_{22})\psi^{(-)}_m
=M_{21}\psi^{(+)}_m+M_{23}\psi^{(+)}_g+M_{24}\psi^{(-)}_g,
\end{equation}
\begin{equation*}
\frac{d\psi^{(+)}_g}{dt}-{\rm i}(\omega_g-{\rm i}M_{33})\psi^{(+)}_g
=M_{31}\psi^{(+)}_m+M_{32}\psi^{(-)}_m+M_{34}\psi^{(-)}_g,
\end{equation*}
\begin{equation*}
\frac{d\psi^{(-)}_g}{dt}+{\rm i}(\omega_g+{\rm i}M_{44})\psi^{(-)}_g
= M_{41}\psi^{(+)}_m+M_{42}\psi^{(-)}_m+M_{43}\psi^{(+)}_g.
\end{equation*}
It is clear that the matrix ${\bf M}$ modifies the mode frequencies
as well as couples the canonical equations for each eigen-variable,
so we appropriately refer to it as a coupling matrix. It involves
the time-derivative of ${\bf C}(t)$ and the matrix ${\bf A}_1$,
which are both proportional to the shear parameter $q$ and to the
azimuthal wavenumber $k_y$. As a result, the coupling matrix
identically vanishes in the shearless limit $q=0$ (rigidly rotating
disc) and/or for axisymmetric perturbations with $k_y=0$, reducing
equations (38) to decoupled one (33). In other words, the
axisymmetric magnetic mode and inertia-gravity waves are not
mutually coupled. So, the linear dynamics of non-axisymmetric
perturbations, involving possible couplings and interactions among
various modes, is richer and much more complex than that of
axisymmetric ones. The homogeneous (left-hand side) parts of
equations (38) describe the individual dynamical behaviour of the
magnetic and inertia-gravity modes. The inhomogeneous (right-hand
side) parts, originating from the non-diagonal components of the
${\bf M}$ matrix, describe mutual couplings between the magnetic
mode bearing the MRI and inertia-gravity waves as well as
interaction between counter-propagating components contained in each
of these modes. It should be stressed that these shear-induced
inhomogeneous terms in the governing equations (responsible for the
linear mode coupling phenomenon) are relatively new in comparison
with related studies on the linear dynamics of perturbations in
magnetized discs
\citep[e.g.,][]{Balbus_Hawley92,Brandenburg_Dintrans06,Johnson07}.
It is seen that compared to the WKBJ limit, the corresponding
instantaneous frequencies/growth rates, $\omega_g$ and $\omega_m$,
of both modes are modified by the diagonal components $M_{11},
M_{22}, M_{33}, M_{44}$. Now the quantities $\omega_m-{\rm
i}M_{11},~\omega_m+{\rm i}M_{22},~\omega_g-{\rm
i}M_{33},~\omega_g+{\rm i}M_{44}$ appear as effective, or modified
instantaneous frequencies/growth rates and determine amplification
properties (instability) of the magnetic and inertia-gravity modes.
Figure 1 shows the squared frequencies/growth rates $\omega^2_m$ and
$\omega^2_g$ as a function of time during the swing of a
perturbation SFH from leading to trailing. $\omega^2_g$ remains
always positive, whereas $\omega^2_m$ becomes negative during a
limited time, indicating the transient non-axisymmetric MRI of the
magnetic mode SFH. Figure 2 shows a similar time history of the
modified frequencies/growth rates, together with $\omega_m$ and
$\omega_g$ for comparison. It is seen that the modified
frequencies/growth rates deviate noticeably from the WKBJ values
$\omega_m$ and $\omega_g$ mostly in the non-WKBJ range
$|k_x(t)/k_z|\lsim 10$ (i.e., $|k_x(t)/k_y|\lsim 5$), implying that
transient effects induced by the shear due to non-axisymmetry play
an important role in the dynamics (instability) of individual modes
at moderate and small radial wavenumbers; at large
$|k_x(t)/k_z|\rightarrow \infty$ the dynamics is mostly adiabatic
and the modified frequencies go to $\omega_m$ and $\omega_g$. Notice
in Fig. 1 that the non-adiabatic interval $|k_x(t)/k_z| \lsim 10$
also covers the range where the MRI would occur (i.e., where
$\omega^2_m<0$) if it were the WKBJ limit, so the growth rate of the
non-axisymmetric MRI is actually determined not by $\omega_m$, as in
the WKBJ and/or axisymmetric cases, but rather by the modified
frequencies $\omega_m\mp {\rm i}M_{11,22}$. Therefore, calculating
the growth rate of the non-axisymmetric MRI using only $\omega_m$,
that is, concentrating on the WKBJ limit during an entire evolution,
as done in several papers \citep[e.g.,][]{Balbus_Hawley92,
Goodman_Xu94, Johnson07}, may not give a full information on the
non-axisymmetric instability of the magnetic mode. This implies that
when studying the dynamics of non-axisymmetric MRI in discs, one
must necessarily take into account the effects of disc flow shear in
full -- first as being responsible for the existence of the MRI
itself (the explicit shear term in linear equations 18-21) and
second as bringing about (implicitly through time-variation of
$k_x$) transient amplification and possible mutual couplings of
different perturbation modes, including the magnetic mode present in
a magnetized disc.

Figure 3 shows the non-diagonal elements of ${\bf M}$ as a function
of time. $M_{12}, M_{21}$ and $M_{34}, M_{43}$ describe the
coupling, respectively, between the counter-propagating
components\footnote{The coupling between the oppositely propagating
components within the same mode is often referred to as
overreflection}, $\psi^{(+)}_m, \psi^{(-)}_m$, of the magnetic mode
and the counter-propagating components $\psi^{(+)}_g, \psi^{(-)}_g$
of inertia-gravity waves. The elements $M_{13}, M_{14}, M_{23},
M_{24}$ and $M_{31}, M_{32}, M_{41}, M_{42}$ describe the coupling
between the magnetic and inertia-gravity modes. In other words,
initially imposed only one either of these two modes can act as a
source for another mode and excite it in the course of evolution. As
expected, all these non-diagonal components reach their maximum
values in the non-adiabatic range $|k_x(t)/k_z|\lsim 10$, where the
effects of the shear due to non-axisymmetry of the modes play a
role. So, the coupling of the magnetic and inertia-gravity modes,
being induced by the disc flow shear, is expected to occur in this
wavenumber range; at large $|k_x(t)|\rightarrow \infty$ the
non-diagonal elements vanish -- there is no energy exchange in the
WKBJ regime. Because of the existence of such a coupling, the
magnetic and inertia-gravity modes actually have strictly separate
identities only in the WKBJ regime, when the characteristic
time-scales (frequencies/growth rates) of these modes are different
(Fig. 1) and the right-hand side terms of equations (38),
responsible for the mode interaction, disappear. By contrast, in the
non-WKBJ regime, the mode time-scales are comparable and the
coupling terms are appreciable (Figs. 2 and 3), so that the modes
cannot be quite disentangled from each other and we have some
mixture of both modes at such times. Finally, we note the following
relations between the various components of the $M$ matrix:
$M_{11}(-k_x)=-M_{22}(k_x)$, $M_{12}(-k_x)=-M_{21}(k_x)$,
$M_{13}(-k_x)=-M^{\ast}_{23}(k_x)$,
$M_{14}(-k_x)=-M^{\ast}_{24}(k_x)$, $M_{33}=M^{\ast}_{44}$,
$M_{31}(-k_x)=-M_{42}(k_x)$, $M_{32}(-k_x)=-M_{41}(k_x)$,
$M_{34}=M^{\ast}_{43}$.

Thus, the shear associated with the disc's differential rotation
plays a twofold role in the dynamics of perturbation modes: 1.
modifies the evolution (stability) of each mode. In particular,
causes the magnetic mode to become magnetorotationally unstable in a
certain range of radial wavenumbers (i.e., we have transient MRI),
2. introduces the phenomenon of linear mode coupling. In the
numerical analysis performed below, we will see the consequences of
this mode coupling in the evolution of the perturbed magnetic field
components.

\section{Dynamics of non-axisymmetric magnetic and inertia-gravity modes}

In this section, we separately investigate the evolution and
amplification properties of non-axisymmetric SFHs (with $k_y\neq 0$)
of basic modes in the system -- inertia-gravity wave and magnetic
mode -- by numerically solving equation (36) (or, equivalently
equations 34 and 35). Then, we concentrate mainly on the magnetic
mode because it determines the MRI in the disc and perform a
comparative analysis of the transient growth of non-axisymmetric and
the exponential amplification of axisymmetric magnetic modes SFHs
during a dynamical time for a range of wavenumbers. For the
numerical integration of equation (36) we used a standard
Runge-Kutta scheme (MATLAB ode45 RK implementation).

We start integration by initially imposing a tightly leading (i.e.,
with $k_x(0)<0, |k_x(0)| \gg |k_y|,|k_z|$) SFH of purely magnetic
mode or purely inertia-gravity waves on the disc flow and trace the
subsequent evolution of the perturbed quantities corresponding to
these initial conditions until $k_x(t)\gg |k_y|,|k_z|$, when the SFH
becomes tightly trailing. At $|k_x(0)|\gg |k_y|,|k_z|$, the SFH
starts out from the WKBJ regime, where the coupling coefficients are
small, so the modes are not coupled with each other and therefore
can be cleanly separated. This, in turn, permits us to isolate
either of these two modes and to insert it as an initial condition
into equation (36). In this adiabatic regime, the eigen-variables
corresponding to counter-propagating components of the magnetic and
inertia-gravity modes evolve independently from each other and are
given by the asymptotic solutions of the homogeneous parts of
equations (38),
\[
\psi^{(+)}_m\approx {\rm e}^{{\rm
i}\int_{0}^{t}\left[\omega_m(t')-{\rm
i}M_{11}(t')\right]dt'},~~~\psi^{(-)}_m\approx{\rm e}^{-{\rm
i}\int_{0}^{t}\left[\omega_m(t')+{\rm i}M_{22}(t')\right]dt'},
\]
\[
\psi^{(+)}_g\approx {\rm e}^{{\rm
i}\int_{0}^{t}\left[\omega_g(t')-{\rm
i}M_{33}(t')\right]dt'},~~~\psi^{(-)}_g\approx{\rm e}^{-{\rm
i}\int_{0}^{t}\left[\omega_g(t')+{\rm i}M_{44}(t')\right]dt'},
\]
To prepare initial conditions to equation (36) corresponding to a
purely magnetic mode imposed at $t=0$, we select out only the
eigen-variable $\psi^{(-)}_m$ from these expressions and set the
initial values of the rest to zero,
$\psi^{(+)}_m(0)=\psi^{(+)}_g(0)=\psi^{(-)}_g(0)=0$. Our choice of
$\psi^{(-)}_m$ as an initial condition is dictated by the fact that
it relates to an exponentially growing branch, i.e., $-{\rm
i}\omega_m=\gamma>0$ when $\omega^2_m=-\gamma^2<0$, where $\gamma$
is the growth rate in the WKBJ regime and/or in the axisymmetric
case. These initial conditions ensure that there is no contribution
from inertia-gravity waves at the beginning. Similarly, to impose
purely inertia-gravity waves in the equations, we select out only
$\psi^{(-)}_g$ and set the initial values of other eigen-variables
to zero, $\psi^{(+)}_g(0)=\psi^{(+)}_m(0)=\psi^{(-)}_m(0)=0$, so
that the magnetic mode is absent in the initial conditions. (Though,
as we checked, for the inertia-gravity waves, the dynamical pictures
corresponding to initially imposed $\psi^{(+)}_g$ or $\psi^{(-)}_g$
display similar properties as far as mode amplification is
concerned.) The magnetic and inertia-gravity modes represent a
complete set of basic modes involved in equations (34) and (35), or
equivalently in equation (36), and any solution can be represented
as their linear combination. So, below we separately analyse the
specific dynamics of magnetic and inertia-gravity modes' SFHs and
compare their transient growth factors, i.e., non-axisymmetric
instability. State vectors ${\bf h}_m$ and ${\bf h}_g$ at $t=0$
associated, respectively, with the initially imposed magnetic and
inertia-gravity modes are constructed as ${\bf h}_m(0)={\bf
C}(0)\cdot[0,\psi^{(-)}_m(0),0,0]^{T}$ and ${\bf h}_g(0)={\bf
C}(0)\cdot[0,0,0,\psi^{(-)}_g(0)]^{T}$.

\begin{figure}
\includegraphics[width=\columnwidth]{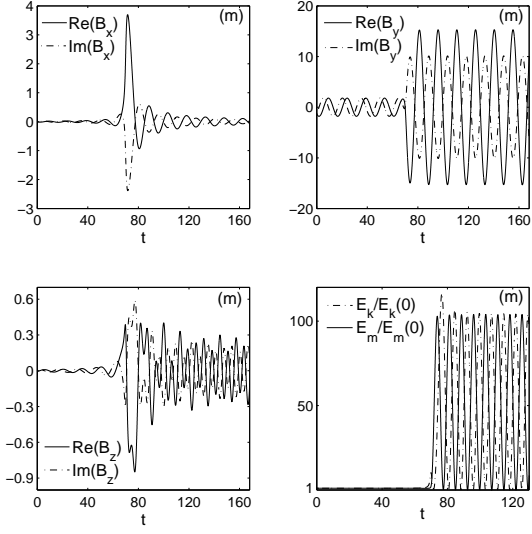} \caption{Evolution of
the magnetic field components pertaining to an initially imposed
magnetic mode SFH with $k_x(0)/k_z=-240, k_y/k_z=2, k_z=6, N^2=0.8$
(solid lines correspond to real parts, dot-dashed to imaginary
parts). The azimuthal field $B_y$ is dominant over the other two
components in the magnetic mode. At the early adiabatic stage of
evolution, $B_y$ oscillates with a constant amplitude, while the
amplitudes of $B_x$ and $B_z$ are small. Then, in the non-adiabatic
region $77<t<83$ (i.e., $|k_x(t)/k_z|<8.1$), all the three
components undergo transient growth and the corresponding kinetic
and magnetic energies amplify by a large factor of 100, indicating
the transient non-axisymmetric MRI exhibited by the magnetic mode.
Afterwards, in the next adiabatic region on the tightly trailing
side at $t > 100$, $B_x$ falls off, while $B_y$ and $B_z$ continue
to oscillate with the frequencies $\omega_m$ and $\omega_g$,
respectively, and with the amplitude of the former being much larger
than that of the latter. In other words, at the non-adiabatic stage,
the dominant azimuthal component of the magnetic field generates
vertical one, which was asymptotically zero at the initial tightly
leading stage. This is a consequence of the linear mode coupling
phenomenon -- generation of the inertia-gravity mode SFH by the
magnetic mode SFH.}
\end{figure}
\begin{figure}
\includegraphics[width=\columnwidth]{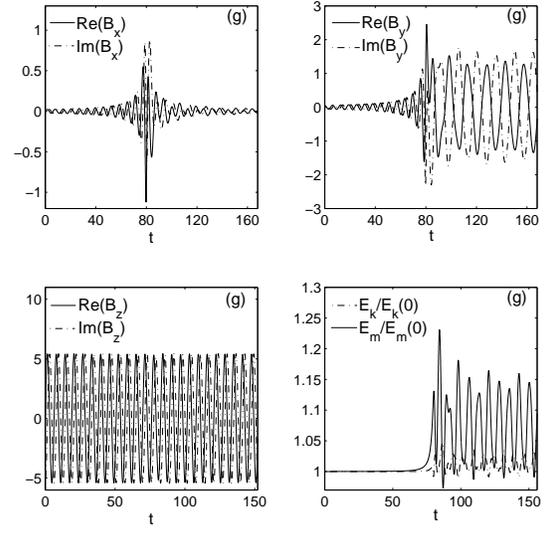} \caption{Evolution
of the magnetic field components pertaining to an initially imposed
inertia-gravity mode SFH with $k_x(0)/k_z=-240, k_y/k_z=2, k_z=6,
N^2=0.8$ (solid lines correspond to real parts, dot-dashed to
imaginary parts). In the inertia-gravity wave SFH, the vertical
field $B_z$ dominates over $B_x$ and $B_y$. Initially, in the
adiabatic region, $B_z$ oscillates with a constat amplitude, while
$B_x$ and $B_y$ are asymptotically zero. Then, during crossing the
non-adiabatic region $77<t<83$ (i.e., $|k_x(t)/k_z|<8.1$), $B_x$ and
$B_y$ components undergo small growth, so that the corresponding
kinetic energy is almost unchanged and the magnetic energy exhibits
a small transient amplification by a factor of 1.25. Afterwards, on
the tightly trailing side at $t > 100$, $B_x$ falls off, while $B_y$
and $B_z$ continue to oscillate with the frequencies $\omega_m$ and
$\omega_g$, respectively. In other words, at the non-adiabatic
stage, the dominant vertical component of the magnetic field
generates azimuthal one, which was asymptotically zero at the
initial tightly leading stage. This is again a consequence of the
linear mode coupling phenomenon -- generation of the magnetic mode
SFH by the inertia-gravity mode SFH.}
\end{figure}

Figures 4 and 5 show the subsequent time-development of the magnetic
field components, $B_x, B_y, B_z$, and normalized by their initial
values kinetic and magnetic energies $E_k/E_k(0), E_m/E_m(0)$ during
the evolution of SFHs corresponding to the above initial conditions
from being tightly leading to tightly trailing. For the initially
imposed magnetic mode SFH (Fig. 4), being still in the WKBJ regime
at $k_x(t)/k_z\ll -1$, the azimuthal magnetic field $B_y$ oscillates
with the Alfv\'{e}n frequency $\omega_m\approx k_z\sqrt{2/\beta}$
and constant amplitude, whereas the radial $B_x$ and vertical $B_z$
components remain much smaller than $B_y$, vanishing in the limit
$k_x(0)\rightarrow -\infty$, that is, the magnetic mode is
characterized by the dominant azimuthal magnetic field perturbation.
This is not surprising if we recall that in a classical Alfv\'{e}n
wave, magnetic field perturbation lies perpendicular to the
background magnetic field and wavevector. Since, in the magnetic
mode, the main restoring force is magnetic tension, the kinetic and
magnetic energies are comparable and larger than the potential
(buoyancy) energy. By contrast, for the imposed inertia-gravity mode
SFH (Fig. 5), at the initial WKBJ stage, $B_z$ oscillates with the
frequency $\omega_g \approx \sqrt{N^2+2k_z^2/\beta}$ and constant
amplitude, whereas $B_x$ and $B_y$ remain much smaller than $B_z$,
vanishing at $k_x(0)\rightarrow -\infty$, that is, in
inertia-gravity waves, dominant is the vertical magnetic field
perturbation. This is due to the fact that the restoring force for
these waves at large $k_x$ is provided mainly by vertical buoyancy
(but is modified by magnetic tension) that induces vertical magnetic
field oscillations via induction equation. As a result, for the
inertia-gravity wave SFH, the kinetic energy is of the same order as
the buoyancy energy and both are much larger the magnetic energy.

As time passes, the radial wavenumber $k_x(t)$ of the SFHs, drifting
along the $k_x-$axis, approaches the non-WKBJ region, where the
elements of the $M$ matrix are appreciable and therefore the
shear-induced effects become important. The components $B_x, B_y$
and $B_z$ for the magnetic mode SFH (Fig. 4) and $B_x$ and $B_y$ for
the inertia-gravity wave SFH (Fig. 5) start to grow transiently.
Then, at about $t_1=77, k_x(t_1)/k_z=-8.1$ the SFHs enter the
non-WKBJ region, cross it and leave at about $t_2=83,
k_x(t_2)/k_z=8.1$ (see Fig.2). For the magnetic mode, the modified
frequency $\omega_m+{\rm i}M_{22}$ associated with the
eigen-variable $\psi^{(-)}_m$ has a positive imaginary part in the
non-WKBJ region (Fig. 2), causing this mode to undergo
transient-exponential amplification in this interval. As a result,
at around $t=80$, all the three components, $B_x, B_y$ and $B_z$,
reach maximum values about 10 times larger than those on entering
the non-adiabatic region, but the azimuthal component, $B_y$, still
remains dominant. Accordingly, the magnetic $E_m$ and kinetic $E_k$
energies of the magnetic mode SFH grow about 100 times during the
transient amplification event. This is essentially the transient MRI
experienced by the non-axisymmetric magnetic mode SFH. On leaving
the non-adiabatic region, in the next adiabatic region, where the
SFH is now tightly trailing (with $k_x(t)/k_z>>1$), $B_x$ gradually
decays and asymptotically, at $k_x(t)\rightarrow \infty$, tends to
zero; $B_y$ continues to oscillate with the amplified amplitude at
an Alfv\'{e}n frequency and the amplitude of $B_z$ levels off at a
constant value, though its oscillation frequency clearly differs
from that of $B_y$ and is equal to the frequency of inertia-gravity
waves, $\omega_g\approx \sqrt{N^2+2k_z^2/\beta}$. This is a direct
consequence of the coupling between the magnetic and inertia-gravity
modes. As noted above, in the inertia-gravity mode SFH dominant is
$B_z$, whereas in the magnetic mode SFH $B_y$ component. In the
initially imposed magnetic mode SFH, the $B_z$ component has been
negligible at the tightly leading stage, because of the absence of
inertia-gravity waves in the initial conditions (see Fig. 4). Then,
in the non-adiabatic region, the non-diagonal coupling elements of
${\bf M}$ matrix become appreciable, resulting in the generation of
the inertia-gravity mode SFH by the original magnetic mode
SFH.\footnote{Note that this coupling occurs in the vicinity of
$k_x(t)\approx 0$, when linear equations do not often have explicit
analytical solutions due to the non-WKBJ nature of this interval and
one has to resort to matching techniques \cite[see
e.g.,][]{Johnson_Gammie05, Heinemann_Papaloizou09a, Pessah_Chan12}.}
So, the $B_z$ component that keeps oscillating with constant
amplitude at large times, in the tightly trailing adiabatic regime
in Fig. 4, corresponds to this newly generated inertia-gravity wave
SFH and hence its oscillation frequency coincides with $\omega_g$.
It can be said that during the evolution of the magnetic mode,
azimuthal magnetic field generates the vertical one due to shear of
the disc's differential rotation. However, the intensity of the
excited inertia-gravity waves is still small compared with that of
the original magnetic mode, whose dominant azimuthal field $B_y$ is
about 50 times larger than $B_z$ of the excited inertia-gravity
wave.

For the initially imposed inertia-gravity wave SFH (Fig. 5), the
effective frequency $\omega_g+{\rm i}M_{44}$ acquires a positive
imaginary part during a short time in the non-WKBJ region because of
the shear-induced component $M_{44}$, however, $B_z$ does not
exhibit any noticeable increase in amplitude and oscillates with
nearly $\omega_g$, during an entire course of evolution of the SFH,
from tightly leading to tightly trailing. Both $B_x$ and $B_y$
undergo moderate transient growth in the vicinity of $t=80$, but
remain smaller than $B_z$. The kinetic energy, $E_k$, shows almost
no growth and the magnetic energy, $E_m$, grows only about 1.5
times. After leaving the non-adiabatic region, when the SFH wraps up
tightly trailing, $B_x$ gradually decays, as in the above case, but
$B_y$ continues to oscillate with a constant amplitude. As seen from
Fig. 5, the oscillation frequency of $B_y$ clearly differs from that
of $B_z$ and is equal to the Alfv\'{e}n frequency, $\omega_m\approx
k_z\sqrt{2/\beta}$. This is again a result of the coupling between
the inertia-gravity and magnetic modes. The initially imposed
inertia-gravity mode SFH with dominant $B_z$ component generates,
during crossing the non-WKBJ region, the magnetic mode SFH with
dominant $B_y$ component. However, the amplitude of the $B_z$
component related to the waves is about 5 times more than that of
$B_y$ belonging to the magnetic mode -- the strength of the
generated magnetic mode SFH is smaller than that of the original
inertia-gravity wave SFH. We obtained a similar result for other
values of $k_y$ and $k_z$.
\begin{figure}
\centering\includegraphics[width=\columnwidth]{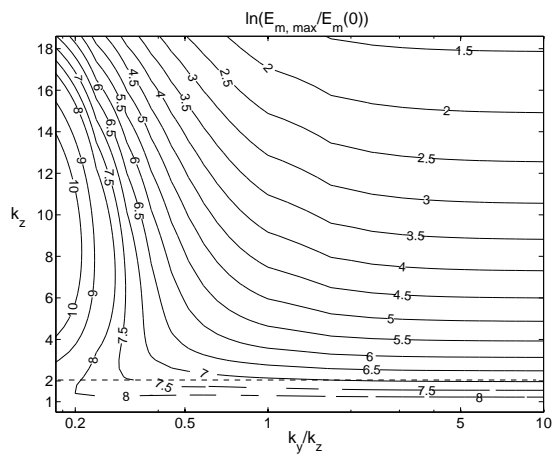}
\caption{Logarithm of the transient growth factor of the magnetic
energy, ${\rm ln}(E_{m,max}/E_{m}(0))$, corresponding to an
initially imposed magnetic mode SFH, as a function of $k_y/k_z$ and
$k_z$ at $N^2=0.8$. The initial energy, $E_m(0)$, is taken at the
tightly leading stage $k_x(0)/k_z\ll -1$. $E_{m,max}$ is the maximum
value achieved by the magnetic energy during the swing of the SFH
from tightly leading to tightly trailing. The values of the growth
factor below $k_z=2$ (dashed lines) are not strictly precise,
because the incompressibility requirement is not valid at such
$k_z$.}
\end{figure}
\begin{figure}
\centering\includegraphics[width=\columnwidth]{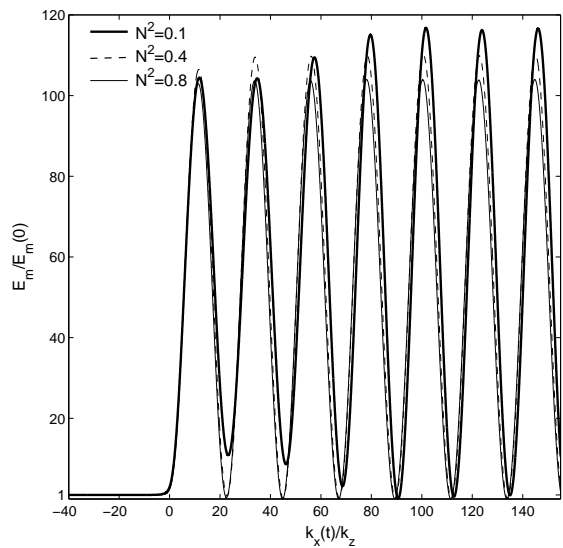}
\caption{Evolution of the magnetic energy corresponding to an
initially imposed magnetic mode SFH with $k_{x}(0)/k_z=-40,
k_y/k_z=2, k_z=6$ versus $k_x(t)/k_z$ at three different
stratifications: $N^2=0.1, 0.4, 0.8$. The evolution of the magnetic
energy is almost independent of the degree of stratification.}
\end{figure}
\begin{figure}
\centering\includegraphics[width=\columnwidth]{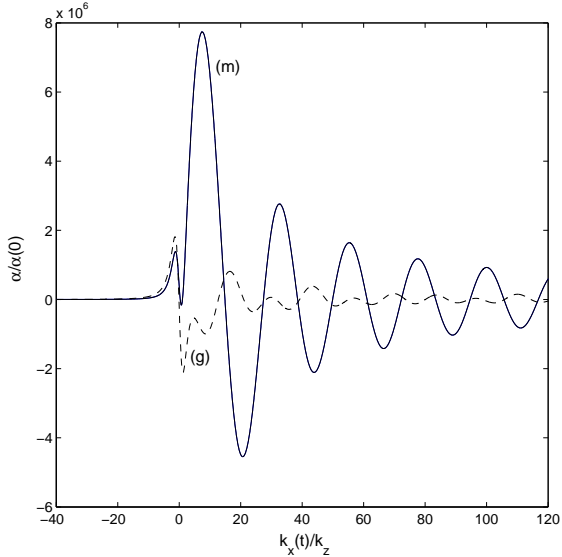}
\caption{Normalized $\alpha/\alpha(0)$ parameters vs. $k_x(t)/k_z$
corresponding to initially imposed magnetic mode (m) and
inertia-gravity wave (g) SFHs at $k_y/k_z=2, k_z=6, N^2=0.8$. The
total stress related to the magnetic mode is positive, while that of
inertia-gravity waves is negative in the dynamically interesting
non-adiabatic region $|k_x(t)/k_z|\lsim 10$ and both decay at late
times.}
\end{figure}

Having separately analysed the SFH evolutions of the magnetic and
inertia-gravity modes, we have found that generally the magnetic
mode tends to grow by orders of magnitude larger factors than the
inertia-gravity waves do, i.e., it is the magnetic mode that
displays the non-axisymmetric MRI and thus remains a dominant driver
of dynamical processes (e.g., angular momentum transport) in
magnetized discs. It should be noted that the amplification of the
magnetic mode is not just transient -- the energy drawn from the
differential rotation during transient amplification phase is not
returned back to the disc, but remains stored in the magnetic mode
which then continues to oscillate with constant amplitude (Fig. 4).
So, below we focus only on the magnetic mode and explore the
efficiency of its growth for various values of the free parameters
of the problem.

To quantify the transient growth of the magnetic mode SFH, we define
the amplification factor -- a ratio of the maximum value achieved by
the magnetic energy over an entire evolution of this SFH from
beginning (tightly leading) to end (tightly trailing) to its initial
value, $E_{m, max}/E_m(0)$. The way we characterize the amount of
amplification, and hence the non-axisymmetric MRI, differs to that
of \cite{Goodman_Xu94}, who used an exponential growth factor
integrated over an entire evolution time,
$\exp(\int_{-\infty}^{\infty}{\rm Im}[\omega_m(t')]dt'$ (in their
notations time changes from $t=-\infty$ to $t=\infty$) to
characterize non-axisymmetric parasitic instabilities afflicting
axisymmetric channel solution. Figure 6 shows the amplification
factor calculated for various $k_y$ and $k_z$. A maximum value at
$k_y/k_z \lsim 0.1$ occurs near $k_z=10$, which approximately
coincides with the critical wavenumber of the most rapidly growing
axisymmetric MRI equal to $k_{z,cr}=(15\beta/32)^{1/2}=13.7\gg 1$
with our normalization. This is easy to understand: at small
$k_y/k_z$, the WKBJ approximation holds during almost an entire
evolution of SFH and one can find the amplification factor by a
simple integration of the WKBJ growth rate given by relation (29)
over the time interval $[t_1,t_2]$, where $t_1$ and $t_2$ are the
moments when drifting $k_x(t)$, respectively, enters and leaves the
unstable region (30), $\exp(\int_{t_1}^{t_2}{\rm
Im}[\omega_m(t')])dt'$. In this case, the extremely large growth
factors ($\sim \exp(10)$) result from the fact that SFHs with small
$k_y/k_z\ll 1$ take long (hundreds of orbits) to cross the unstable
region. In this respect, the growth of non-axisymmetric SFHs with
such small $k_y$, despite being huge, do not actually play a role
because it occurs over times much larger than the dynamical/orbital
time. At $k_y/k_z \gsim 1$, the growth factor is independent of
$k_y/k_z$ and increases with decreasing $k_z$ (see also Fig. 10).
This indicates that as opposed to the (nearly) axisymmetric mode,
which has its maximum exponential growth rate at larger vertical
wavenumbers, most of growth of the non-axisymmetric magnetic mode
occurs preferably at smaller $k_z \lsim 5$. However, the very large
($\sim\exp(8)$) values of the growth factor below $k_z\lsim 2$
(dashed lines in Fig. 6) are not really precise, since they are
related to SFHs with vertical wavelengths comparable to the disc
scale-height for which compressibility effects, neglected here
because of the Boussinesq approach, are important (see also next
subsection, Fig. 10).

We also analysed the role of vertical stratification in the dynamics
of non-axisymmetric modes. It is well-known that (convectively
stable) stratification does not affect the fastest growing
axisymmetric perturbations \citep{Balbus_Hawley91}. In Fig. 7, we
show the time-development of the normalized magnetic energy,
$E_m/E_m(0)$, for the magnetic mode SFH at various stratifications
at fixed $k_y$ and $k_z$. It is seen that the growth factor of
energy weakly depends on the stratification, especially during the
exponential-transient amplification phase around $k_x(t)=0$. This is
easily explained: the magnetic mode's amplification is mainly
provided by the combined action of rotation and magnetic tension,
with buoyancy playing only a minor role. At later times, at the
tightly trailing stage ($k_x(t)/k_z=2k_x(t)/k_y \gg 1$), the
energies oscillate with amplitudes that are a bit larger at smaller
stratifications. Thus, stratification has a weak influence on the
growth of the non-axisymmetric MRI in the linear regime, however, it
plays an important role in the non-linear development and saturation
of the MRI \citep{Davis_etal10}.

In Fig. 8 we also plot the time-development of the total stress, a
sum of Reynolds and Maxwell stresses, given by equation (22),
corresponding to the initially imposed leading harmonics (with
$k_x(0)/k_y<0$) of inertia-gravity waves and magnetic mode. Being
still at the leading stage ($k_x(t)/k_y<0$), both are positive and
comparable to each other, but at the trailing stage, at
$0<k_x(t)/k_z=2k_x(t)/k_y \lsim 10$, transport due to the magnetic
mode is positive and larger by absolute value than the negative
transport due to inertia-gravity waves. So, inertia-gravity wave
harmonics in this range of wavenumbers, like convection, tend to
transport angular momentum inwards \citep[see also][]{Balbus03}.
However, the harmonics of both modes oscillate at large times, in
the tightly trailing regime, and their respective $\alpha$, being
also oscillatory, gradually decay. So, in the linear theory, the
sign of transport cannot be determined with certainty. In the
non-linear regime, however, inertia-gravity waves may amount to
moderate fraction of or become even comparable to the magnetic mode,
despite the prevalence of the latter in the linear regime. In this
case, both these dynamically intertwined modes (not only the
magnetic mode) are to be taken into account when studying a
nonlinear (turbulent) state in incompressible, stratified and
magnetized discs. Hence, understanding the true nature of turbulent
angular momentum transfer in such discs requires a scrupulous
analysis of the contributions of magnetic mode and inertia-gravity
waves in the transport.

Recently, \citet{Pessah_Chan12} have carried out an analogous study
of the linear dynamics of non-axisymmetric shearing MHD waves, but
with $k_z=0$, in an incompressible unstratified shearing box
threaded by a weak magnetic field. Although they focused on
accretion disc boundary layers with negative shear, $q<0$, and
therefore not subject to the standard MRI, their results regarding
the evolution of energy and total stress for a magnetic mode SFH
resemble those described above. So, we would like to draw a parallel
between the findings presented in that paper and ours. In the
shearing box without vertical stratification, for 2D perturbations
with $k_z=0$, only the magnetic mode -- Alfv\'{e}n wave -- survives,
whereas inertia-gravity wave disappears. The transient amplification
of these Alfv\'{e}n waves, which was investigated by Pessah \& Chan,
has non-exponential character and can be attributed to the
shear/non-normality of the boundary layer flow. \footnote{See also
\cite{Chagelishvili_etal97b} for shear-induced transient
amplification of MHD waves in magnetized plane shear flows.} As a
result, the corresponding amplification factors in the non-adiabatic
region, in the vicinity of $k_x(t)=0$, are smaller than those of the
magnetic mode SFHs in our case with the MRI, as shown in Figs. 4 and
6; outside the non-adiabatic region, the magnetic mode SFH behaves
like that of 2D Alfv\'{e}n wave does. Specifically, the evolution of
the total stress for the magnetic mode SFH shown in Fig. 8 and for
2D Alfv\'{e}n wave SFH are qualitatively similar: both experience
transient growth near $k_x(t)=0$, however, the one related to the 2D
Alfv\'{e}n wave is negative there due to the negative shear, and
oscillate around zero with decreasing amplitude at large
$|k_x(t)/k_y|\gg 1$. Also, the SFH energies, starting with a
constant value, after a transient amplification phase, do not decay
and keep on oscillating about a time-independent value at late
times. In our 3D case with stratification, inertia-gravity waves
also come into play. Although their transient amplification is small
compared with that of the magnetic mode, as demonstrated above, they
couple with the magnetic mode in the non-adiabatic regime near
$k_x(t)=0$ and must be taken into account for a correct/fuller
description of the mode dynamics.

\subsection{comparative analysis of non-axisymmetric and axisymmetric SFHs' growth}

An analysis of the competition between the non-axisymmetric and
axisymmetric modes' linear dynamics is of primary interest, because
this competition will determine the main cause/ingredient of the
transition to turbulence and the resulting non-linear turbulent
state. If the initial amplitude of perturbations is very small,
axisymmetric magnetic mode, characterized by permanent exponential
growth with constant rate, eventually wins in this competition and
forms an axisymmetric channel flow
\citep{Hawley_etal95,Bodo_etal08,Longaretti_Lesur10}. However, if
the initial amplitudes of both non-axisymmetric and axisymmetric
modes are sufficiently large, then outcome is not so obvious. Below
we examine possible dynamics in the latter case.

It is well-known that the most unstable axisymmetric magnetic mode
SFH has the growth rate $\gamma_{max}=0.75\Omega$, i.e., the
dynamical (e-folding) timescale $t_c=1/\gamma_{max}=1.33\Omega^{-1}$
\citep{Balbus_Hawley91,Goodman_Xu94}. This is the shortest
characteristic timescale of the axisymmetric MRI and effectively a
dynamical time too, as it is of the same order as the shear time
$(q\Omega)^{-1}=0.67\Omega^{-1}$. So, a natural choice for the time
interval during which the non-axisymmetric and axisymmetric growths
are to be compared is therefore $t_c$.

We have seen above that the magnetic energy of the non-axisymmetric
magnetic mode SFH, evolving from far tightly leading ($k_x(0)/k_z\ll
-1$) to tightly trailing phases ($k_x(t)/k_z\gg 1$), increases by
orders of magnitude (see Fig. 6). The amplification factor of the
energy increases with decreasing the azimuthal wavenumber, $k_y$. At
small $k_y$, as noted above, the amplification factor is determined
to a good approximation by the WKBJ growth rate given by expression
(29) and the corresponding duration of this amplification is the
time the radial wavenumber, $k_x(t)$, takes to cross the unstable
interval (30) in the wavenumber space. However, this swing time can
be quite long in comparison with the dynamical/e-folding timescale,
$t_c$, and therefore the most powerful growth occurring for such
small $k_y$ does not really play a decisive role as far as disc's
dynamical instability is concerned. In some sense, ``truly''
non-axisymmetric dynamical growth, that occurs over times of order
$t_c$, starts from about $k_y/k_z\gsim 1$ and $k_z\gsim 1$. Thus, to
analyze the relative importance of the transient growth of
non-axisymmetric magnetic mode SFHs compared with the fastest
growing axisymmetric one during $t_c$ and to understand their role
in the system dynamics, we adopt the concept of ``optimal''
perturbations \citep{Butler_Farrell92, Farrell_Ioannou96,
Farrell_Ioannou00,Bakas_etal01}, which is widely employed by
geophysical/meteorological community. As applied to our problem, the
strategy of the optimal perturbation approach is to consider the
evolution of a single magnetic mode SFH with some initial $k_x(0),
k_y$ and $k_z$ within the time interval $[0, t_c]$, to calculate the
resulting transient amplification and then to explore how it changes
with varying these wavenumbers. Specifically, at each point in the
$(k_x(0), k_y, k_z)$ space, at the initial moment of time, $t=0$, we
prepare initial conditions corresponding to a purely magnetic mode
SFH (i.e., initially impose $\psi_m^{(-)}(0) \neq 0 $ and
$\psi_g^{(+)}(0)=\psi_g^{(-)}(0)=\psi_m^{(+)}(0)=0$) and trace
subsequent evolution during the time $t_c$. From Figs. 4 and 7 we
see that the energy of a magnetic mode SFH undergoes most of its
growth on the trailing side at $k_x(t)/k_z \lsim 10$ and,
afterwards, turns into oscillatory with constant amplitude.
Obviously, optimal SFHs are expected to be in this range (see
below). Each non-axisymmetric SFH drifts in the ($k_x,k_y$)-plane
along the $k_x$-axis with speed $qk_y$, so that the optimal SFH with
the initial radial wavenumber $k_x(0)$ will have the wavenumber
$k_x(t_c)=k_x(0)+qk_yt_c$ at $t_c$. Hence, we can regard the energy
of the optimal perturbation as a function of $k_x(t_c)$, $k_y$ and
$k_z$.

We define the transient amplification factor of the magnetic mode
SFH in terms of the associated magnetic energy, $E_m(t)$, as the
ratio of its final, at $t=t_c$, to the initial, at $t=0$, values.
Since this is to be compared with the largest exponential growth
factor of the magnetic energy of the most unstable axisymetric
magnetic mode during the same time interval, that is, with
$\exp{(2\gamma_{max}t_c)}=\exp(2)$, we introduce a new parameter --
a relative amplification factor
\[
f_e(k_x(t_c),k_y,k_z)=\exp(-2)\frac{E_m(k_x(t_c),k_y,k_z)}{E_m(k_x(0),k_y,k_z)}.
\]
This perameter conveniently measures the non-axisymmetric growth of
the magnetic energy against axisymmetric one during the dynamical
time and thus characterizes relative significance of
non-axisymmetric magnetic mode SFHs with respect to axisymmetric
ones. SFHs with those wavenumbers that yield the maximum (or close
to maximum) value of $f_e$ represent optimal (or close-to-optimal)
perturbations.

\begin{figure}
\centering\includegraphics[width=\columnwidth]{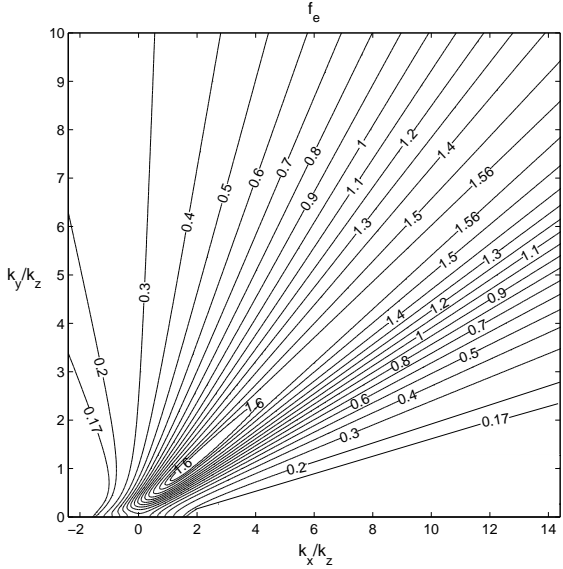}
\caption{Isocontours of the relative amplification factor
$f_e(k_x(t_c),k_y,k_z)$ in the $(k_x/k_z,k_y/k_z)$-plane at $k_z=7$
and $N^2=0.8$. It achieves an extremum (maximum), $f_{e,max}=1.56$,
on the trailing side ($k_x/k_z>0$), on the line $k_y=0.57k_x$. This
extremum is almost constant all along this line, reaching a slightly
larger peak, or optimal value $f_{e,opt}=1.65$ at the lower end
$k_x/k_z=1.8, k_y/k_z=1.03$. This peak corresponds to the optimal,
while the values $f_e>1$ near the line of extremum, to
close-to-optimal SFHs. As is clear, the amplification factors of the
non-axisymmetric magnetic mode SFHs can be comparable to or even
larger than that of the maximally growing axisymmetric one during
the characteristic time of the latter.}
\end{figure}

\begin{figure}
\centering\includegraphics[width=\columnwidth]{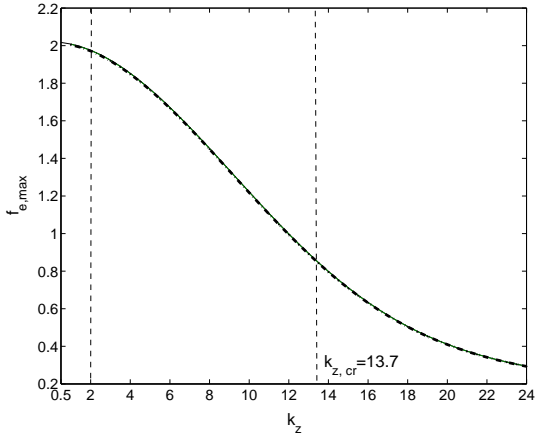}
\caption{Extremum value, $f_{e,max}$, as a function of $k_z$ at
$N^2=0.8$ (solid line) and $N^2=0.01$ (dash-dotted line). (We also
tried $N^2=0.1,0.4$, but this curve is practically insensitive to
stratification). The first vertical dashed line serves as a border
of the compressibility regime $k_z\lsim 2$. $f_{e,max}$ increases
with decreasing $k_z$, reaching the largest values near the
compressibility margin. For reference, the second vertical dashed
line marks the critical wavenumber $k_{z,cr}=13.7$ of the fastest
growing axisymmetric MRI.}
\end{figure}

Figure 9 shows isocontours of the relative amplification factor,
$f_e$, in the ($k_x/k_z,k_y/k_z$)-plane at $k_z=7$ and $N^2=0.8$. In
this map, the values range from 0.17 to the optimal (largest) value
$f_{e,opt}=1.65$ at $k_x/k_z=1.8, k_y/k_z=1.03$. The values above
unity $f_e$ reaches in the vicinity of the line $k_y=0.57k_x$, which
can be labeled as the line of extremum (maximum) of $f_e$
(obviously, the optimal SFH also lies on this line). Farther up
along the line of extremum, it quickly levels off to an
asymptotically constant value $f_{e,max}=1.56$. All the magnetic
mode SFHs with $f_e \gsim 1$ are referred to as ``close-to-optimal''
harmonics. These harmonics are situated in the vicinity of the
extremum line and located on the trailing side of the
($k_x,k_y$)-plane. Thus, it is evident that there is a fairly broad
range of $k_x(t_c),k_y$ for which the energy of non-axisymmetric
magnetic mode SFHs can transiently grow by factors larger than a
purely exponential amplification of energy for the most unstable
axisymmetric magnetic mode during the dynamical time $t_c$. This
does clearly demonstrate that non-axisymmetric magnetic mode
perturbations can prevail over, or compete with axisymmetric ones
during $t_c$ at those wavenumbers. Contrasting Figs. 6 and 9, both
describing non-axisymmetric amplification of the magnetic mode, we
see different trends with respect to $k_y$ at fixed $k_z$. In Fig.
6, where amplification factor, being calculated over much larger
time interval for the magnetic mode SFH evolution, increases with
decreasing $k_y$, because SFHs with smaller $k_y$ stay much longer
in the unstable region (30). By contrast, in the second case in Fig.
9, the evolution time is fixed to $t_c$, which is not enough for
nearly axisymmetric ($k_y\ll 1$) harmonics with various initial
$k_x(0)$ to undergo considerable growth, whereas non-axisymmetric
ones with moderate $k_y/k_z\gsim 1$ manage to get amplified more
within the same time.

These findings should be important for further development in the
non-linear stage and, ultimately, for turbulence (and possibly
coherent structures) that may emerge in this system. Of course, if
the amplitudes of initial perturbations are very small and the
linear regime holds for a long time, the axisymmetric/exponential
MRI will eventually dominate. Optimal non-axisymmetric perturbations
undergo most of growth only during the dynamical time interval.
However, if the optimal and close-to-optimal SFHs have sufficiently
large initial amplitudes, the non-linearity will come into play
during the dynamical time and the further exponential growth of the
axisymmetric MRI (e.g., axisymmetric channel flow) may not persist.
In other words, in such a course of events, the optimal and
close-to-optimal non-axisymmetric SFHs can prevail over axisymmetric
SFHs and determine the onset of a non-linear regime and its
dynamics, including angular momentum transport parameter $\alpha$.

As is well-known, the exponential growth rate of an axisymmetric
magnetic mode SFH reaches a maximum at the critical wavenumbers
$k_{x,cr}=0, k_{z,cr}=(15\beta/32)^{1/2}$, decreases at
$k_z<k_{z,cr}$ and $k_z>k_{z,cr}$ and reaches zero at $k_z=0$ and
$k_z=\sqrt{q\beta}$ \citep{Balbus_Hawley91}. Let us now examine how
the non-axisymmetric relative amplification factor (found above at
given $k_z$) varies with $k_z$. Figure 10 shows the maximum,
$f_{e,max}$, along the above-mentioned line of extremum as a
function of $k_z$. Note that $f_{e,opt}$ is close to $f_{e,max}$, so
we have chosen to plot only the latter, because as evident from Fig.
9, those close-to-optimal SFHs with values of $f_e$ near to
$f_{e,max}$ generally occupy larger area in the wavenumber space
than the optimal SFH. From this figure it is seen that the relative
amplification factor of non-axisymmetric magnetic mode SFH
monotonically increases with decreasing $k_z$. For $k_z<12$,
$f_{e,max}>1$ indicating that at such vertical wavenumbers the
transient growth of non-axisymmetric magnetic mode perturbations
always exceeds that of the most unstable axisymmetric ones during
the dynamical time. Therefore, non-axisymmetric MRI becomes dominant
over, or at least competitive with the axisymmetric one in the disc
dynamics. Note that at $k_z\lsim 2$, when vertical wavelength of
perturbations becomes comparable to the disc scale-height,
compressibility is at work and the part of this curve at such $k_z$,
calculated here in the incompressible limit, does not give quite
reliable values of the non-axisymmetric amplification factor.
Nevertheless, Fig. 10 correctly predicts a trend -- the
amplification of non-axisymmetric magnetic mode SFH is the larger,
the smaller $k_z$ is. By contrast, as mentioned above, the
axisymmetric MRI attains its largest growth rate at critical
vertical wavelengths a few times smaller than the scale-height,
$\lambda_{z,cr}/H=2\pi/Hk_{z,cr}=2\pi(32/15\beta)^{1/2}=0.46$.
Extrapolating this to the compressible case, non-axisymmetric
magnetic mode perturbations, which transiently grow the most, are
expected to be more influenced by compressibility than axisymmetric
ones. \emph{So, in order to accurately characterize the
non-axisymmetric MRI in discs with net vertical magnetic field, one
should take into account compressibility.} This may be related to
the saturation of magnetic energy and stresses in MRI-driven
turbulence that exhibits dependence on compressibility and hence the
latter plays an important role in the non-linear development of the
non-axisymmetric MRI with non-zero net magnetic flux
\citep[e.g.,][]{Sano_etal04, Pessah_etal07, Guan_etal09}.

\section{Summary and discussion}

In this paper, we have investigated the linear dynamics of
non-axisymmetric perturbations in a vertically stratified Keplerian
disc with a weak uniform vertical magnetic field. We mainly
considered incompressible perturbations, but, in the final part,
speculated on the dynamics in the compressible case. An analysis was
performed in the framework of the local shearing box approximation.
As is usually done in this case, perturbations were decomposed into
spatial fourier harmonics (SFHs), or shearing plane waves, and
substituted into linearized ideal MHD equations of the shearing box.
The temporal evolution of the amplitudes of SFHs was followed by
numerical integration in time of the resulting set of linear
differential equations. In contrast to previous linear studies on
the non-axisymmetric MRI in discs
\citep[e.g.,][]{Balbus_Hawley92,Tagger_etal92,Brandenburg_Dintrans06,Johnson07},
we took a different approach and rewrote these equations in the
canonical form for characteristic variables, or eigen-variables
(equations 38) that allowed us to identify and describe all the
modes/types of perturbations contained in this disc model. In this
respect, our analysis offers a deeper insight into the specific
dynamics and amplification properties of individual modes, allows us
to easily capture a new effect of their mutual couplings brought
about by the disc's differential rotation. There are two basic modes
in the considered incompressible disc model: inertia-gravity waves
and magnetic mode. The magnetic mode, which is driven mainly by
magnetic tension force but modified by rotation, displays the MRI.
Inertia-gravity waves are driven by (convectively stable) vertical
buoyancy and rotation and are spectrally stable, though
non-axisymmetric ones can still undergo transient growth at small
radial wavenumbers due to disc flow shear. For this reason, we
focused primarily on non-axisymmetric modes, since the dynamics of
axisymmetric modes in magnetized discs is quite well understood
\cite[e.g.,][]{Balbus_Hawley91,Ogilvie98}. Besides, shear effects
manifest themselves only for non-axisymmetric perturbations and
hence one can say that the dynamics of non-axisymmetric modes is
more diverse and richer than that of axisymmetric ones. We compared
the transient growth of initially imposed non-axisymmetric
inertia-gravity waves and magnetic mode SFHs during evolution from
tightly leading to tightly trailing and found that the latter
exhibits much larger transient growth factors. Also, the stresses
associated with it, dominate those due to inertia-gravity waves. In
other words, the magnetic mode is a primary mode taking part in
energetic processes and determines the MRI in discs. Another
important and relatively new effect we observed is the coupling of
magnetic mode and inertia-gravity waves. The evolution of an
initially imposed magnetic mode SFH has shown that, for a certain
range of vertical and azimuthal wavenumbers, the transient
amplification of the magnetic mode SFH is accompanied by the
generation of inertia-gravity wave SFHs, though with smaller
amplitude -- the non-axisymmetric MRI acts as a new source of
inertia-gravity waves. This is a consequence of the linear mode
coupling phenomenon resulting from disc flow shear/non-normality
\citep{Chagelishvili_etal97b}. It demonstrates that, although
inertia-gravity waves are not a main participant in energetic
processes defining disc instability, still their presence should be
taken into account, together with the more important magnetic mode,
in order to form a complete dynamical picture of the
non-axisymmetric MRI in stratified discs. In magnetized discs,
inertia-gravity waves carry predominantly vertical magnetic field
perturbation, whereas in the dominant magnetic mode its mostly
azimuthal. The linear coupling between these two modes implies that
vertical magnetic field can be generated by azimuthal one and vice
versa due to the disc flow shear.

Another goal was a parallel analysis of the dynamics of
non-axisymmetric and axisymmetric magnetic modes. The accepted
course of events in the presence of non-zero net vertical magnetic
field is the following: initially the linear axisymmetric MRI arises
whose subsequent non-linear development leads to the formation of a
channel solution. Afterwards, on this channel flow, the secondary
instabilities of (parasitic) non-axisymmetric perturbations develop
\citep{Goodman_Xu94, Latter_etal09, Pessah_Goodman09}. Typically, in
the stability analysis of a channel flow, one considers small
non-axisymmetric perturbations to the channel solution once it has
been well formed, i.e., reached amplitudes much larger than the
background magnetic field \citep{Goodman_Xu94, Latter_etal09}. In
this case, the growth rate of parasitic instabilities turns out to
be much larger than that of the channel itself. This fact
considerably simplifies the analysis, because the background shear
and rotation can be neglected and the channel can be assumed steady.
In the above papers, a case of not very large channel amplitudes
(compared to the background magnetic field) is also studied.
Stability of this (intermediate amplitude) channel solution against
non-axisymmetric perturbations is complicated by shear, rotation and
the growth (non-steadiness) of the channel itself. This problem is
attacked using the technique of shearing waves \citep{Goodman_Xu94}.
But what the above papers have in common is that they all consider
the linear stability of already well-developed channel flows being
subject to non-axisymmetric (parasitic) perturbations with
amplitudes smaller than that of the channel itself.

The above dynamical picture can be dramatically altered if there is
a competition between axisymmetric and transient non-axisymmetric
growths from the early linear stage of evolution during the
characteristic $e$-folding time, $t_c$, of the most unstable
axisymmetric MRI (since a typical time for the development of the
non-linear channel is of the order of $t_c$). If the initial
amplitude of non-axisymmetric perturbations is not small, their
transient growth comes into play and these perturbations may become
a key ingredient of the overall dynamics. We compared the
amplification factor of the magnetic energy of non-axisymmetric
magnetic mode SFHs with various radial, azimuthal and vertical
wavenumbers with that of the most unstable axisymmetric magnetic
mode SFH during $t_c$. From this comparative analysis it turned out
that at vertical wavelength larger than about the critical
wavelength of the axisymmetric MRI (i.e., for $k_z < k_{z,cr}$, see
Fig. 10), the growth factor of non-axisymmetric magnetic mode SFHs
is larger than that of the most unstable axisymmetric ones. In other
words, the transient growth of these optimal and close-to-optimal
non-axisymmetric SFHs exceeds the maximal exponential growth of
axisymmetric SFHs over the dynamical time. When forming a complete
dynamical picture, one should also take into account the fact that
the set of the close-to-optimal SFHs is substantially larger than
the set of the exponentially growing axisymmetric ones, as evident
from Fig. 9. As a result, axisymmetric SFHs can be overwhelmed by
the close-to-optimal SFHs if they have sufficiently large initial
amplitudes. In this case, non-linear interaction between
non-axisymmetric and axisymmetric SFHs will start just after
transient amplification, leading directly to turbulent state,
without/before the formation of the channel flow. In other words,
non-axisymmetric magnetic perturbations can prevent the development
of the channel in its early growth stage. This course of events may
be another reason for the absence of channels in some simulations of
the MRI-driven turbulence with non-zero net vertical magnetic flux
configurations \citep[e.g.,][]{Hawley_etal95,Bodo_etal08}.

Our investigation also demonstrated that amplification factor of
non-axisymetric magnetic mode SFHs increases with decreasing
vertical wavenumber and reaches its maximal value at vertical and
azimuthal wavelengths comparable to the disc scale-height (see Figs.
6 and 10). In weakly magnetized discs, the axisymmetric MRI is
essentially incompressible, as the largest growth rate comes at
wavelength smaller than the disc scale-height. By contrast,
according to our incompressible calculations, the largest values of
the non-axisymmetric amplification factor are achieved at vertical
wavelengths of perturbations comparable to the disc scale-height for
which compressible wave modes become inevitably involved (via mode
linear coupling) in the dynamics \citep{Tevzadze_etal03}.
Consequently, for a realistic analysis of the non-axisymmetric MRI
in discs, one should take into account compressibility. In support
of this, we would like to mention that the saturation level of
MRI-driven turbulence, involving non-axisymmetric Fourier harmonics,
in numerical simulations with box sizes of the order of scale-height
does depend on compressibility
\citep{Sano_etal04,Pessah_etal07,Guan_etal09}. In addition, our
preliminary analysis indicates that the linear coupling of spiral
density waves and non-axisymmetric MRI occurs in the compressible
case. Due to this coupling, non-axisymmetric magnetic mode with
azimuthal and vertical wavelength comparable to the disc
scale-height can effectively excite spiral density waves. This is of
the same nature as the generation of high-frequency density waves by
aperiodic (non-oscillatory) modes in non-magnetized discs
\citep{Bodo_etal05,Mamatsashvili_Chagelishvili07,Mamatsashvili_Rice11}
and in MRI-driven turbulence \citep{Heinemann_Papaloizou09b}.
However, in the latter paper, a set up with zero net magnetic flux
was considered, where the linear MRI was apparently absent (or at
least was weak) and density waves were primarily generated only by
the vortical mode. The situation can be different in the case of
non-zero net vertical magnetic flux we considered, where the main
mode generating density waves is not the vortical but the magnetic
mode.

\section*{Acknowledgments}
GRM, GB and PR acknowledge the financial support from the Instituto
Nazionale di Astrofisica (INAF) under the project CRA 1.06.11.80. We
would like to thank the anonymous referee for helpful comments and
suggestions that improved the presentation of our work.

\bibliographystyle{mn2e}
\bibliography{biblio}

\end{document}